\documentclass[twocolumn,dvipsnames]{aastex63}
\usepackage{amsmath}

\submitjournal{ApJ}

\shorttitle{HYPER}
\shortauthors{He et al.}

\graphicspath{{./}{figures/}}

\begin{document}

\title{A Hydro-Particle-Mesh Code for Efficient and Rapid Simulations of the Intracluster Medium}

\correspondingauthor{Yizhou He}
\email{yhe2@andrew.cmu.edu}

\author{Yizhou He}
\affil{McWilliams Center for Cosmology, Department of Physics, Carnegie Mellon University, Pittsburgh, PA 15213, USA}

\author{Hy Trac}
\affil{McWilliams Center for Cosmology, Department of Physics, Carnegie Mellon University, Pittsburgh, PA 15213, USA}
\affil{NSF AI Planning Institute for Physics of the Future, Carnegie Mellon University, Pittsburgh, PA 15213, USA}

\author{Nickolay Y. Gnedin}
\affil{Theory Department; Fermi National Accelerator Laboratory; Batavia, IL 60510, USA}
\affil{Kavli Institute for Cosmological Physics; The University of Chicago; Chicago, IL 60637, USA}
\affil{Department of Astronomy \& Astrophysics; The University of Chicago; Chicago, IL 60637, USA}

\begin{abstract}
We introduce the cosmological HYPER code based on an innovative hydro-particle-mesh (HPM) algorithm for efficient and rapid simulations of gas and dark matter. For the HPM algorithm, we update the approach of \citet{1998MNRAS.296...44G} to expand the scope of its application from the lower-density intergalactic medium (IGM) to the higher-density intracluster medium (ICM). While the original algorithm tracks only one effective particle species, the updated version separately tracks the gas and dark matter particles as they do not exactly trace each other on small scales. For the approximate hydrodynamics solver, the pressure term in the gas equations of motion is calculated using robust physical models. In particular, we use a dark matter halo model, ICM pressure profile, and IGM temperature-density relation, all of which can be systematically varied for parameter-space studies. We show that the HYPER simulation results are in good agreement with the halo model expectations for the density, temperature, and pressure radial profiles. Simulated galaxy cluster scaling relations for Sunyaev-Zel'dovich (SZ) and X-ray observables are also in good agreement with mean predictions, with scatter comparable to that found in hydrodynamic simulations. HYPER also produces lightcone catalogs of dark matter halos and full-sky tomographic maps of the lensing convergence, SZ effect, and X-ray emission. These simulation products are useful for testing data analysis pipelines, generating training data for machine learning, understanding selection and systematic effects, and for interpreting astrophysical and cosmological constraints.
\end{abstract}


\section{Introduction}\label{sec:intro}

Cosmological simulations have greatly improved our understanding of the formation and evolution of the cosmic structures throughout our universe and are widely used to interpret observations and design new instruments and surveys. N-body simulations of gravitational dynamics make detailed and reliable predictions for the distributions of dark matter, which forms the backbone of structure formation, and dark energy responsible for the accelerated expansion of the universe \citep[e.g.][]{2005Natur.435..629S,2009MNRAS.398.1150B,2011ApJ...740..102K,2012MNRAS.426.2046A,2014arXiv1407.2600S,2016MNRAS.457.4340K,2021MNRAS.tmp.1529I}. Hydrodynamic simulations, which model the coupled evolution of dark matter and cosmic gas \citep[e.g.][]{2012MNRAS.427.1024I,2014Natur.509..177V,2015MNRAS.450.1349K,2015MNRAS.446..521S,2016MNRAS.455.2778F,2016MNRAS.463.1797D,2016MNRAS.463.3948D,2017MNRAS.467.4739K,2017MNRAS.465.2936M,2017MNRAS.465..213B,2017MNRAS.470.1121T,2018MNRAS.473.4077P,2019MNRAS.486.2827D}, are also able to predict many directly observable properties of cosmic gas and galaxies.

Understanding the large scale structure of our universe requires two parts: (i) an accurate solution to the equations of motion for the dark matter and (ii) physically reasonable approximations for the behavior of baryonic components of the universe. Together, these two components build the foundation of the modern observational cosmology. Dark matter N-body simulations have achieved significant
progress in developing the understanding of the structure of dark matter halos \citep[e.g.][]{1996ApJ...462..563N,1997ApJ...490..493N,2004MNRAS.349.1039N} and their clustering \citep[e.g.][]{2006Natur.440.1137S,2008ApJ...688..709T,2013MNRAS.433.1230W,2016MNRAS.456.2361B}. They were instrumental in establishing the $\Lambda$CDM cosmological model as the dominant paradigm for the nature of both dark matter and dark energy, but suffer from the fundamental limitation of being incapable of providing any direct prediction for the baryonic component. Thus, the hydrodynamic simulations that also simulate the baryons that form the visible components in our universe are crucial for interpretation and calibration of the results of observational data. A detailed review of the modern hydrodynamic simulations studying the properties, growth and evolution of galaxies is given in \citet{2020NatRP...2...42V}.

Hydrodynamic simulations are generally preferred for solving the non-linear physics of structure formation and predicting the survey observable dependence on cosmological parameters. However, the effective volumes of modern surveys keep growing, and  achieving the science goals of these surveys requires numerical simulations of exceptionally large volumes - both for correctly capturing the statistics of the rare objects and for computing the covariance matrices between the observables. Simulations in spatial volumes comparable to the surveys in size are generally too expensive to make many large-scale mock observations and explore both astrophysical and cosmological parameter space. In the face of increasing demand for multiple realizations of simulated mock catalogs for comparison with the large-scale structure observations, fast approximate approaches for dark matter simulations based on semi-numerical methods and linear perturbation theory have been developed. For example, PTHALOS \citep{2002MNRAS.329..629S} has been used for efficiently generating mock galaxy distributions, PINOCCHIO \citep{2002MNRAS.331..587M} is capable of accurately predicting formation and evolution of individual dark matter halos, COLA \citep{2013JCAP...06..036T} and FastPM \citep{2016MNRAS.463.2273F} can be used for cheaply generating large ensembles of accurate mocks that properly account for non-linear evolution. These fast approximate methods have shown their ability to reduce computational complexity and required computational resources by orders of magnitude without sacrificing accuracy on large scales. 

Though we have seen significant progress in various approaches aiming to speed up dark matter only N-body simulations, there is still a notable lack of fast approximate hydro simulation methods. Previously, \citet{1998MNRAS.296...44G} used the particle-mesh (PM) solver for dark matter dynamics and allowed for the additional gas pressure force to approximate hydrodynamics. Their hydro-particle-mesh (HPM) algorithm substantially relies on the existence of a tight temperate-density relation in the intergalactic medium (IGM) and has been successfully used to model the high-redshift Lyman alpha forest with moderate precision \citep{2002ApJ...580...42M}. However, the tight correlation between the gas density and temperature in the low density IGM breaks down in denser regions. Yet, it is possible to extend the range of validity of HPM-like techniques further. For example, in order to model the ICM of galaxy clusters we can adopt empirical or simulated ICM pressure profiles \citep[e.g.][]{ 2010A&A...517A..92A, 2012ApJ...758...75B,2021ApJ...908...91H} and build a mapping relation between the gas temperature or pressure and some properties of cosmic gas that can be captured by, say, a simple PM solver. Such an approach will allow to implement a fast approximate method for modeling hydrodynamics in the high-density ICM; the gas physics can then be modeled very efficiently in both the IGM and the ICM regime, which together fill most of the spatial volume in a fast hydro simulation. 

This paper introduces an innovative hydro-particle-mesh code for efficient and rapid (HYPER) simulations of gas and dark matter. HPM simulations take approximately two to three times as long to run as PM simulations and are orders of magnitude faster than expensive hydrodynamic simulations. HYPER allows one to systematically vary the ICM and IGM models to study different baryonic physics and effects. We organize this paper as follows: Section \ref{sec:models} discusses the model for radial profiles of dark matter and gas in the ICM. Section \ref{sec:methods} briefly reviews the HPM algorithm in the IGM, then describes the implementation of the HPM for the ICM regime, including how to modify the PM code to calculate the designed HPM fields and infer the gas temperature and pressure from the local field information using a pre-computed mapping derived from a given ICM model. In Section \ref{sec:results} we evaluate our new fast hydro simulation performance by comparing the radial profiles, the integrated halo quantities, and the statistical quantities of the tSZ effect in our simulation to the predictions of the ICM model we use to implement the HPM algorithm. In Section~\ref{sec:conclusion} we conclude with our findings for the output of a HYPER simulation, and at the end, we also bring out some perspective of use cases and the future extensions of this work.

\clearpage
\section{Models} \label{sec:models}

This section introduces our approach for constructing an analytical model for the radial profiles of different components in galaxy clusters, under the assumption of hydrostatic equilibrium. It is based on thoroughly studied modeling work on dark matter distribution and gas pressure support. We also briefly introduce the analytical relation used by the HPM algorithm \citep{1998MNRAS.296...44G} to simulate gas thermal properties in the IGM.

\subsection{Halo Model}

As a model for the mass distribution in a dark matter halo of a galaxy cluster we adopt a universal NFW density profile \citep{1997ApJ...490..493N}
\begin{equation}\label{eq:rho_m}
\rho_{\rm m}(r) \simeq \frac{\rho_\mathrm{s}}{\frac{r}{r_\mathrm{s}}(1+\frac{r}{r_\mathrm{s}})^2} ,
\end{equation}
 where $r_\mathrm{s}$ is the scale radius and $\rho_\mathrm{s}$ is the scale density. These two variables can be specified by the halo concentration as
\begin{equation}
    r_\mathrm{s} = R_\mathrm{500c}/c_\mathrm{500c}
\end{equation}
and
\begin{equation}
    \rho_\mathrm{s}=\frac{500\rho_{\mathrm{crit}}(z)c^3_\mathrm{500c}}{3\left[\mathrm{ln}(1+c_\mathrm{500c})-c_\mathrm{500c}/(1+c_\mathrm{500c})\right]},
\end{equation}
where $R_\mathrm{500c}$ defines the radius of a spherical overdensity region with average density 500 times of the critical density $\rho_\mathrm{crit}(z)$. The total mass of the overdensity region can be used to define the halo mass as $M_\mathrm{500c}=\frac{4\pi}{3}500\rho_{\mathrm{crit}(z)}R_\mathrm{500c}^3$. The halo concentration $c_\mathrm{500c}$ has been calibrated as a function of cluster mass and redshift \citep[e.g.][]{2015ApJ...799..108D}. Thus given the halo mass and redshift, we can specify the halo density profile of dark matter. The total mass enclosed within the radius $r$ then has an analytical expression
\begin{equation}
\begin{split}
    M(r)&=\int_0^r4\pi x^2\rho_{\rm m}(x)dx\\
    &= 4\pi\rho_\mathrm{s}r_\mathrm{s}^3\left[\ln\left(\frac{r+r_\mathrm{s}}{r_\mathrm{s}}\right)-\frac{r}{r+r_\mathrm{s}}\right].
\end{split}
\end{equation}

We use the NFW profile to approximate the total matter distribution, considering that dark matter dominates the mass contribution to the halo. The error on the total mass profile is mainly caused by the difference between the gas density profile and the NFW profile, which is generally considered to be a goof approximation to the dark matter distribution of halos. Though the deviation of the normalized gas density profile from the NFW model may not be negligible, the difference between the total matter density of halos and the NFW profile is much less significant since gas is subdominant in the total matter distribution. The difference between the total matter density profile and the NFW model decreases by a factor of the baryonic fraction $f_\mathrm{b}\sim0.15$ compare to the gas density profile and should be much less than unity. Thus a lot of analytical work choose to model the total matter density profile with the NFW model, so that $\rho_{\rm m}(r)=\rho_\mathrm{dm}(r)+\rho_\mathrm{gas}(r)\simeq \rho_\mathrm{NFW}(r)$ \citep[e.g.][]{2012MNRAS.423.1534O,2014MNRAS.442..521S}. X-ray and lensing studies have also shown that NFW profiles can generally provide adequate descriptions of the mass distribution of cluster halos \citep[e.g.][]{2006MNRAS.372..758M,2007MNRAS.379..209S,2012MNRAS.421.3147M}.

\subsection{ICM Model}\label{subsec:icmmodel}

We construct an analytical model of the ICM based on the assumption of hydrostatic equilibrium and models for describing the gas pressure of galaxy clusters. The gas thermal pressure profile has been widely studied by both hydrodynamic simulations and SZ and X-ray observations. In most of these studies, the thermal pressure of gas generally follows an analytical form
\begin{equation}\label{eq: pressure model}
    P_\mathrm{th}(M,r)=P_\mathrm{500c}(M,z)\mathbb{P}(r/R_\mathrm{500c})f(M)g(z),
\end{equation}
where $P_\mathrm{500c}$ is the pressure scale derived from the self similar scaling relation \begin{eqnarray}
    P_\mathrm{500c}&=1.65\times10^{-3}E(z)^{8/3} \nonumber \\
    &\times\left(\frac{M_\mathrm{500c}}{3\times 10^{14}M_\odot}\right)^{2/3},
\end{eqnarray}
where $E(z)=H(z)/H_0$ and $\mathbb{P}(x=r/R_\mathrm{500c})$ is a generalized NFW (GNFW) model \citep[e.g.][]{2007ApJ...668....1N},
\begin{equation}
    \mathbb{P}(x)=\frac{P_0}{(c_{500}x)^\gamma[1+(c_\mathrm{500}x)^\alpha]^{\frac{\beta-\gamma}{\alpha}}}.
\end{equation}
Parameters $P_0,c_{500},\alpha,\beta,\gamma$ are normally fitted to the observed or simulated scaled pressure profiles.

The ICM gas pressure profile modelled with only the GNFW term assumes a simple self-similar behavior \citep[e.g.][]{1986MNRAS.222..323K,2005RvMP...77..207V}. However, several numerical and observational works \citep[e.g.][]{2010A&A...517A..92A,2017MNRAS.469.3069G} claimed to observe a deviation from the self-similar scaling relation in the gas pressure, which we denote by the additional terms $f(M)$, $g(z)$ in Eq.~\ref{eq: pressure model}. We adopt power-law forms for these terms, $f(M)=(M/M_\star)^{\alpha_\mathrm{P}}$ and $g(z)=E(z)^{c_\mathrm{P}}$, where $M_\star$ is a chosen constant referred to as the characteristic cluster mass. However, the dependence on halo mass and redshift of the gas pressure model needs further study, as the studies on these deviation terms with numerical simulations and observations have not reached the final agreement.

In this work, we adopt the debiased pressure profile (DPP) from a recent study of the gas pressure model of the galaxy clusters \citep{2021ApJ...908...91H}, which adjusted the Universal Pressure Profile (UPP) \citep{2010A&A...517A..92A} for hydrostatic mass bias in X-ray observation. The DPP model strictly follows the GNFW model and get rid of the terms describe the deviation from $f(M)$ and $g(z)$ for the gas thermal pressure $P_\mathrm{th}(r)$. The GNFW parameters $P_0,c_{500},\alpha,\beta,\gamma$ of the pressure model are [5.048,1.217,1.192,5.490,0.433].

Further studies of the cluster outskirts with the latest cosmological simulations \citep[e.g.][]{0004-637X-725-2-1452,2012ApJ...751..121N,2012ApJ...758...74B,2013ApJ...777..151L,0004-637X-792-1-25,2017MNRAS.469.3069G} and the observations of galaxy clusters \citep[e.g.][]{2009PASJ...61.1117B,2009MNRAS.395..657G,2009A&A...501..899R,2010PASJ...62..371H,2010ApJ...714..423K,doi:10.1111/j.1365-2966.2011.18526.x,2011Sci...331.1576S} showed existence of the non-thermal pressure support, which is non-negligible particularly in the outskirt of galaxy clusters. The non-thermal pressure is mainly due to the non-thermal gas processes like virialized bulk motions and turbulent gas flows, which are primarily generated by mergers and accretion during the cluster formation. Then the total pressure of gas $P_\mathrm{tot}(r)$ should consist of both thermal and non-thermal contributions, modeled by
\begin{equation}
\begin{split}
    P_\mathrm{tot}(r) &= P_\mathrm{th}(r)+P_\mathrm{nth}(r)\\
    &= f_\mathrm{th}P_\mathrm{tot}(r) + f_\mathrm{nth}P_\mathrm{tot}(r)
\end{split}
\end{equation}
where $f_\mathrm{th}(r)$ and $f_\mathrm{nth}(r)$ are ratios of the thermal and nonthermal terms to the total pressure.

Studying the sample of 65 galaxy clusters from a high resolution hydrodynamic cosmological simulation, \citet{0004-637X-792-1-25} characterized the non-thermal pressure fraction profile $f_\mathrm{nth}(r)$, and found it was universal across redshift and cluster mass of the studied 65 galaxy clusters. When scaling the cluster radii with respect to the mean matter density of the universe, the thermal fraction we mentioned above can be then expressed as  
\begin{equation}
    f_\mathrm{th}(r)= 1-f_\mathrm{nth}(r)=A\left\{1+\exp\left[-\left(\frac{r/R_\mathrm{200m}}{B}\right)^{\gamma}\right]\right\},
\end{equation}
with the best fitted parameters $A=0.452$, $B=0.841$, $\gamma=1.628$.

Then based on the assumption of the cluster dynamical state being in hydrostatic equilibrium,
\begin{equation}
    \frac{{\rm{d}} P_\mathrm{tot}(r)}{{\rm{d}}r}=-\rho_\mathrm{gas}(r)\frac{GM(r)}{r^2},
\end{equation}
we are able to predict the gas density profile from analytic models for the enclosed mass function of halo $M(r)$, the gas thermal pressure profile $P_\mathrm{th}(r)$ and the gas thermal fraction $f_\mathrm{th}(r)$ as
\begin{equation}
    \rho_\mathrm{gas}(r)=-\frac{r}{GM(r)}\frac{f_\mathrm{th}(r)\frac{{\rm{d}}P_\mathrm{th}(r)}{{\rm{d}}r}-P_\mathrm{th}(r)\frac{{\rm{d}}f_\mathrm{th}(r)}{{\rm{d}}r}}{f_\mathrm{th}^2(r)} .
\end{equation}
The gas temperature profile of the cluster is derived by the ideal gas equation of state,
\begin{equation}
    T_\mathrm{gas}(r)=\frac{\mu P_\mathrm{th}(r)}{k_\mathrm{B}\rho_\mathrm{gas}(r)},
\end{equation}
where $\mu$ is the mean mass per gas particle and $k_\mathrm{B}$ is the Boltzmann constant.

We emphasize that the analytical models we choose for both the thermal and non-thermal pressure of the gas in this paper are only presented as an example, not some hard-coded choice. In HYPER, the hydro part that drives the gas particles hydrodynamics can be constructed based on general models for the gas pressure. Thus we can systematically vary the pressure profiles and study different ICM models with HYPER.  

\subsection{IGM Model}\label{subsec:IGM}

In the low-density IGM regime, where the local density is less than the mean density of the universe that $\rho_{\rm m}/\bar{\rho}_{\rm m}\lesssim 10$, shock-heating is not important, and the gas physics is set by different processes than in the ICM regime in \citet{1998MNRAS.296...44G}.

\citet{1997MNRAS.292...27H} proposed a semi-analytical method to predict the temperature-density relation for any given cosmology and reionization history, using the Zel'dovich approximation. They find a tight correlation (to better than 10\%) between the gas density and temperature (and hence pressure as well), well described by a power-law,
\begin{equation}
    T =  T_0 \Delta^{\gamma-1},
\end{equation}
where $T_0(z)$ is a function of redshift only and is of order of $10^4$K, $\Delta$ is the relative gas density, $\Delta=\rho_{\rm gas}/\bar{\rho}_{\rm gas}$,
and $\gamma$ is between 1 and 1.62.

Both $T_0$ and $\gamma$ are found evolving with time. \citet{1997MNRAS.292...27H} derived analytical approximations to the temperature-density relation in a scenario when the universe reionizes rapidly, which is often considered a suitable approximation for low redshift evolution of $\gamma$ and $T_0$. 

\clearpage
\section{Methods} \label{sec:methods}
\begin{figure*}[th]
\includegraphics[width=\textwidth]{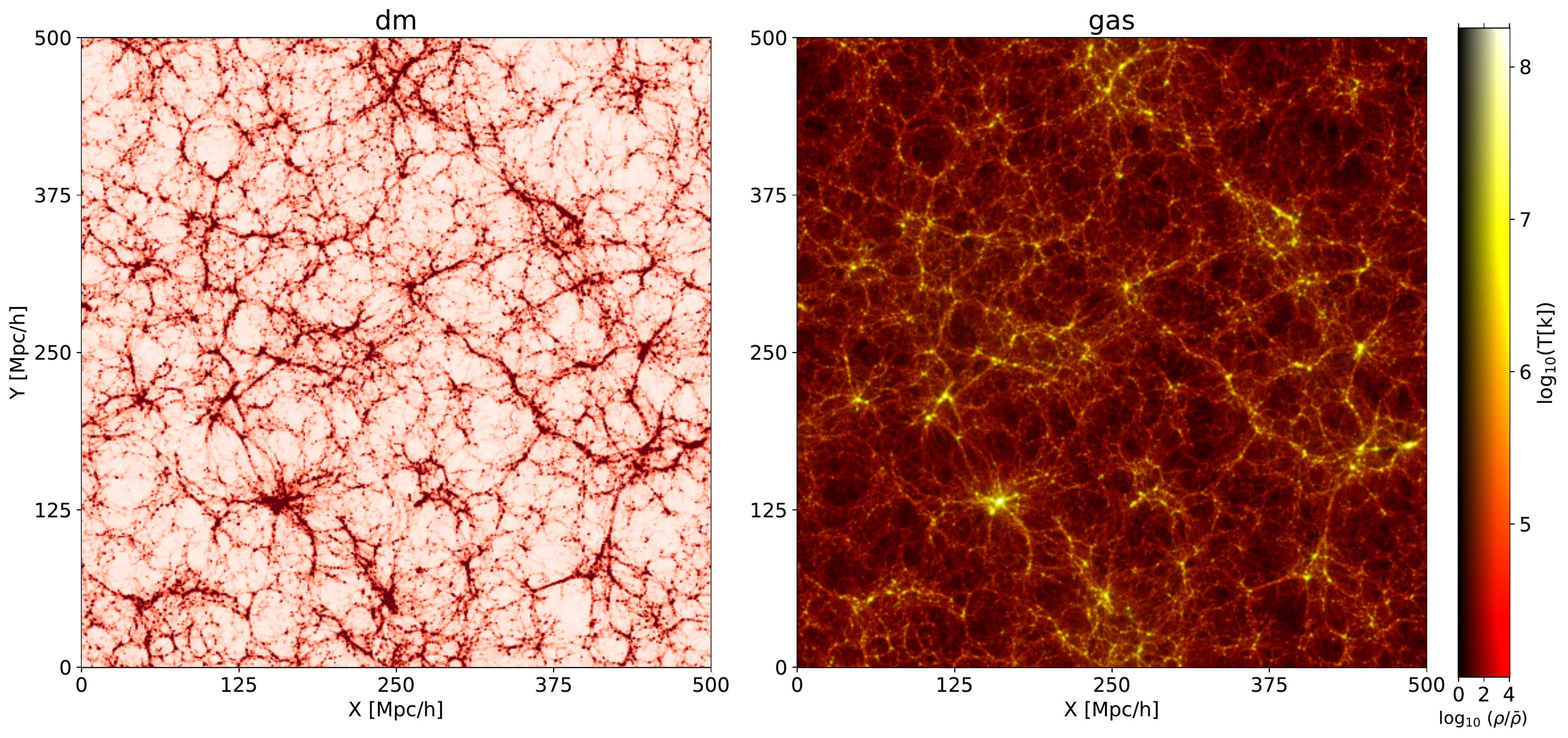}
\caption{Visualization of dark matter and gas in the HYPER simulation at $z = 0.0$. The simulation box size is 500 Mpc/h
per side with a thickness of 20 Mpc/h. {\bf Left:} Shown is a projection of the dark matter density field. The cosmic large-scale structures like massive collapsed halos, elongated filaments and near-empty voids in the HYPER simulation can be easily identified in the thin slice. {\bf Right:} The projection of gas density field is shown to have a strong correlation with the dark matter distribution. In this slice, brightness indicates the projected mass density and color hue visualizes the mean projected temperature (dull-red to brilliant-yellow indicating cold to hot, as shown by the color bar aside).}\label{fig:sim_vsl}
\end{figure*}

In this section we introduce our further development of the HPM algorithm, which has already shown its potential for being an efficient and accurate alternative to full hydrodynamic simulations for simulating the low column density Lyman-alpha forest \citep{1998MNRAS.296...44G}, by expanding it from the IGM regime to the high-density ICM with the help of the analytical ICM model discussed in Section \ref{sec:models}. 

\subsection{Particle-Mesh}

The PM solver, which laid the foundation of the HPM algorithm, is widely used for tracking the evolution of collisionless dark matter \citep[e.g.][]{1981csup.book.....H}. We start with introducing a PM code developed by \citet{2004NewA....9..443T}, which demonstrates how we resolve the evolution of collisionless dark matter particles in our HPM algorithm. This PM code has been adapted for solving Newton's equations of motion for dark matter particles in a Friedman-Robertson-Walker (FRW) universe. In the expanding FRW background, we use comoving coordinates $\vec{x}_{\rm c}=\vec{x}/a$, where the scale factor $a$ is governed by the Friedman equation
\begin{equation}
    \frac{\mathrm{d}a}{\mathrm{d}t} = aH_0(\Omega_\mathrm{m}a^{-3}+\Omega_\Lambda)^{1/2},
\end{equation}
assuming a spatially flat background. And to preserve the time-invariant conservation form of the Euler fluid equations, we take a new variable
\begin{equation}
    \mathrm{d}\tau=\frac{\mathrm{d}t}{a^2},
\end{equation}
for the time coordinate \citep[e.g.][]{1980MNRAS.192..321D,1995ApJS...97..231G,1998MNRAS.297..467M, 1998ApJS..115...19P, 2004NewA....9..443T}.

The time-dependence of the cosmological expansion remains only in the gravitational source term, 
\begin{equation}
    \nabla^2\Phi=4\pi aG(\rho_{\rm m}-\bar\rho_{\rm m}).
\end{equation}
Here the matter density is per comoving volume $\mathrm{d}^3\vec{x}_{\rm c}=a^{-3}\mathrm{d}^3\vec{x}$, and is related to the proper mass density by $\rho_{\rm m} = a^3\rho_{\rm m}^\mathrm{p}$. The dynamical equations of dark matter particle under the spatial and time coordinate transformations are
\begin{align}
    &\frac{\mathrm{d}\vec{x}_{\rm c}}{\mathrm{d}\tau} = \vec{v},\\
    &\frac{\mathrm{d}\vec{v}}{\mathrm{d}\tau}=-\vec{\nabla}\Phi,
\end{align}
where code quantities velocity $\vec{v}$ and gravitational potential $\Phi$ are related to the corresponding physical proper quantities by
$\vec v = a\vec v^\mathrm{p}$, $\Phi = a^2\Phi^\mathrm{p}$.

The PM solver solves the equation of motion for the dark matter particles treating the gravitational force as a field and approximating it on a uniform mesh. We firstly use the "Cloud-in-a-Cell" (CIC) scheme \citep{1981csup.book.....H} to deposit  mass, including both the dark matter and gas, to the mesh to create the density field, then calculate the gravitational potential $\Phi$ by solving the Poisson's equation, with the Fast Fourier Transform. In Fourier space, the modified Poisson's equation of the continuous system is expressed as
\begin{equation}
    \tilde{\Phi}(\vec{k})= \frac{4\pi aG\tilde{\rho}_{\rm m}(\vec{k})}{k^2},
\end{equation}
For the PM solver in our HPM algorithm, the Green function is obtained from directly transforming the finite-difference approximation of the Laplacian in the Poisson's equation,
\begin{equation}
    \tilde{\Phi}(\vec{k})= \frac{4\pi aG\tilde{\rho}_{\rm m}(\vec{k})(\frac{\Delta l}{2})^2}{\sin^2(\frac{\Delta lk_x}{2})+\sin^2(\frac{\Delta lk_y}{2})+\sin^2(\frac{\Delta lk_z}{2})},
\end{equation}
where $\Delta l = L/N$ is the length of unit grid cell, $L$ is the length of simulation box, and $N$ is the number of mesh cells per side. When calculating the force field, differential operators, such as the gradient $\nabla$, are replaced by the finite difference approximations. Potential and accelerations at particle positions are obtained by interpolating on the array of mesh-defined values.

We adopt the $kick\text{-}drift\text{-}kick$ (KDK) leapfrog integrator also used for GADGET-2 \citep{2005MNRAS.364.1105S} in our simulation, updating the position and velocity for each particle with the time evolution operator $U(\Delta\tau)=K(\frac{\Delta\tau}{2})D(\Delta\tau)K(\frac{\Delta\tau}{2})$. The kick and drift operators, K and D, are defined as
\begin{align}
      K(\Delta\tau)=\left\{
\begin{array}{ll}
      &\vec{x}_i \rightarrow \vec{x}_i \\
      &\vec{v}_i \rightarrow \vec{v}_i+\vec{a}_i\Delta\tau\\
\end{array} 
\right.\\
D(\Delta\tau)=\left\{
\begin{array}{ll}
      &\vec{x}_i \rightarrow \vec{x}_i+\vec{v}_i\Delta\tau\\
      &\vec{v}_i \rightarrow \vec{v}_i\\
\end{array} 
\right.
\end{align}
where the acceleration $\vec{a}_i$ is interpolated for each particle from the acceleration field on the grid.

\subsection{Hydro-Particle-Mesh}\label{subsec:hpm}
In a hydro simulation, the main difference between dark matter and gas, dynamically, is that the latter is subject to pressure forces in addition to the gravity. Our HPM method is designed by modifying the dark matter only PM algorithm to also solve the dynamical equation for gas particles,
\begin{equation}\label{eq:gas_motion}
    \frac{{\rm{d}}\vec{v}_\mathrm{gas}}{{\rm{d}}\tau}=-\nabla\Phi-\frac{\nabla P}{\rho_\mathrm{gas}},
\end{equation}
which, in addition to the gravitational force, also includes the pressure force $-\nabla P/\rho_\mathrm{gas}$. 
The comoving gas pressure $P$ 
is related to the proper pressure $P_\mathrm{p}$ by $P = a^5 P_\mathrm{p}$.

The pressure force on gas particles is inferred from HPM variables calculated and saved for each gas particle interpolated to the particle positions on HPM fields and a mapping relation derived from the gas model of ICM and IGM we have already discussed in Section \ref{subsec:icmmodel} and \ref{subsec:IGM}. We will show how to select the HPM fields in Section \ref{subsec:hpmvar} and how to construct the mapping relationship in detail in Section \ref{subsec:hpmtable}. We calculate the gravitational force $-\nabla\Phi$ following the same procedure as for the original PM code described above. The pressure force $-\nabla P/\rho_\mathrm{gas}$ 
is calculated by depositing the particle pressure
to the mesh, applying the finite difference, and finally interpolating it back to the particles in the simulation. Specific details of the implementation of the gas pressure calculation with the constructed mapping relation between the gas particle thermal properties and the HPM variables is discussed in Section \ref{subsec:hpmtable}. 

Notice that we are no longer adopting a single component model to reduce computational cost as in the original HPM. As explained by \citet{1998MNRAS.296...44G}, in the IGM regime the gas pressure is dynamically subdominant on large scales; hence, if one is only interested in modeling the baryonic component, one can treat the part of the gravitational force acting on baryons from the dark matter as if the dark matter followed the baryons. However, in the high-density ICM regime, pressure and gravity are comparable quantities, dark matter distribution can not be accurately approximated by the distribution of baryons, and the difference between the two components can not be neglected when solving for the evolution of baryons. Hence, this difference between the IGM and the ICM requires one to track gas and dark matter particles separately. 
  
The results in this paper are drawn from a HPM simulation of box whose per side length $L = 500h^{-1}$Mpc with periodic boundary conditions and equal numbers of dark matter and gas particles $N_\mathrm{dm}=N_\mathrm{gas}= 1024^3$, and the mesh size is set to be $N_\mathrm{mesh}=4096^3$. The simulation runs from start point $z=100$ to $z=0$ while the hydro part for gas particle is turned on at $z=6$ takes $\sim$ 4500 CPU hours in total. In the simulation, we adopt a “concordance” $\Lambda$CDM model ($\Omega_\mathrm{m}=0.3$, $\Omega_{\Lambda}=0.7$, $\Omega_\mathrm{b}=0.045$, $h=0.7$ and $\sigma_8=0.8$).

\subsection{HPM Variables and Fields}\label{subsec:hpmvar}

For the ICM model we have discussed in Section \ref{subsec:icmmodel}, if the halo mass $M$ and the distance of the gas to the halo center $r$ are specified, one can derive the gas thermal properties from these two variables,
\begin{equation}
    X = X(M,r),
\end{equation}
where $X$ refers to the gas thermal quantities like temperature $T$ or pressure $P$.
However, in the HPM algorithm, the halo mass and the displacement of gas particles with respective to the halo center are not directly available (without adding on-the-fly halo finding), and evaluating the gas properties in the ICM region after specifying the halo it resides in will also enforce unwanted spherical symmetry. Our goal is to avoid mapping the gas temperature or pressure directly from the halo mass and the gas particle displacement in the halo. Instead of identifying the halo where gas resides and locating the gas particle with respective to the halo center, we choose to use designed HPM variables and fields that could more accurately reflect the local environment information of simulated particles. The HPM variables interpolated to the position of gas particles with the HPM fields should also show enough connection to the halo mass and radius when assuming an ideal spherical symmetry scheme, so that we can use them to build a pre-computed mapping for inference of the gas thermal properties based on the ICM model. Since it requires two variables $M$ and $r$ to calculate the thermal quantities of the gas in the ICM model, the number of HPM fields used for calculating the HPM variables as alternative to halo variables $M$ and $r$ should also be at least 2 to break the degeneracy.

According to the description above, the HPM fields chosen should satisfy two conditions to keep our HPM algorithm efficient enough: (a) the calculation of the HPM fields in the HPM method should be efficient enough, and (b) derivation of HPM variables in the adopted ICM model should be simple, preferably with analytical expressions as a function of halo mass and radius.

In HYPER, we adopt the matter density and a new field, the scalar force, to be our HPM variables for the inference of gas thermal properties. We choose the matter density because this quantity is readily available, since it is used to solve the Poisson's equation and is saved for each particle in the simulation. For the other HPM variable, the scalar force, we create a new variable based on the idea that originates from the gravitational force calculation. Recall, that the gravitational force (per unit mass) is defined by
\begin{equation}
    \vec{f}_\mathrm{g}(\vec{x})=\int \frac{G\rho_{\rm m}(\vec{x}^\prime)(\vec{x}^\prime-\vec{x})}{|\vec{x}^\prime-\vec{x}|^3}{\rm d}^3\vec{x}^\prime=G\rho_{\rm m}(\vec{x})\otimes \frac{\vec{x}}{|\vec{x}|^3},
\end{equation}
where $G$ is the gravitational constant, $\otimes$ denotes the operation of convolution. For the newly designed scalar force (per unit mass) variable we simply replace $\frac{\vec{x}}{|\vec{x}|^3}$ in the convolution with $\frac{1}{|\vec{x}|^2}$. The new variable shares the same units as the gravitational force, but is a scalar instead,
\begin{equation}
    f_\mathrm{scalar}(\vec{x})=G\rho_{\rm m}(\vec{x})\otimes \frac{1}{|\vec{x}|^2}.
\end{equation}
The scalar force $f_\mathrm{scalar}$ satisfies the requirements for the HPM variables. Thus we can quickly implement it in the PM solver, since the Fourier transform of $\frac{1}{|\vec{x}|^2}$ is 
\begin{equation}
    \int \frac{1}{|\vec{x}|^2}e^{i\vec{x}\cdot\vec{k}}{\rm d}^3\vec{x}= \frac{2\pi^2}{k}
\end{equation}
and
\begin{equation}
    \tilde{f}_\mathrm{scalar}(\vec{k})= \frac{2\pi^2G\tilde{\rho}_{\rm m}(\vec{k})}{k},
\end{equation}
which can be solved directly analogous to computation of the gravitational potential with the PM solver  described in Section~\ref{subsec:hpm}.
And in the halo model, we can calculate the radial profile of the scalar force by using the NFW profile for the matter density of the halo convolved with $\frac{1}{|\vec{x}|^2}$. For the scalar force of a halo at radius $|\vec{x}|=r$, we can can rewrite the integral in the spherical coordinates, so that
\begin{equation}\label{eq:f_scalar}
\begin{split}
    f_\mathrm{scalar}(r)&=
    \int G\rho_{\mathrm{NFW}}(\vec{x}^\prime)\times\frac{1}{|\vec{x}-\vec{x}^\prime|^2}{\rm d}^3\vec{x}^\prime\\
    &=\int_{0}^{\infty}G\rho_{\mathrm{NFW}}(r^\prime)\int_0^\pi\frac{2\pi (r^\prime)^2{\rm d}\theta}{(r^\prime)^2+x^2-2r^\prime x\cos\theta}{\rm d}r^\prime\\
    &=\frac{2\pi G r_\mathrm s}{x}\left[\int_{0}^r\frac{\rho_\mathrm{s}}{(1+r^\prime/r_\mathrm{s})^2}\ln\left(\frac{r^\prime+r}{r-r^\prime}\right){\rm d}r^\prime+\right.\\
    &\ \ \ \ \left.\int_{r}^\infty\frac{\rho_\mathrm{s}}{(1+r^\prime/r_\mathrm{s})^2}\ln\left(\frac{r^\prime+r}{r^\prime-r}\right){\rm d}r^\prime\right]\\
    &=\frac{2\pi G\rho_\mathrm{s}r_\mathrm{s}\ln(r/r_\mathrm{s})}{1-(r/r_\mathrm{s})^2}.
\end{split}
\end{equation}
Thus, for the given halo mass and radius we can calculate the scalar force with Eq.~\ref{eq:f_scalar}, which will facilitate our construction of the mapping relation between the HPM variables and the gas thermal properties in Section~\ref{subsec:hpmtable}.

Other candidates for the HPM variables that have been considered are gas density or gravitational acceleration, since they are also computed by HPM code. However, if the gas density is one of the fields determining the gas temperature and pressure, then there is no way to prevent artificial numerical fragmentation when the local Jeans' length becomes too small \citep{1997ApJ...489L.179T} with the fixed spatial resolution of the uniform PM grid. With using the total mass density as the HPM variable, such numerical artifacts are greatly suppressed.

For another candidate, the gravitational acceleration mentioned above is similar to the matter density since dark matter also dominates the source of gravitational force. Thus, this quantity is also very stable against numerical fragmentation. However, the gravitational acceleration is significantly affected by finite mesh resolution and is underestimated in the center regions of simulated halos. The finite resolution effect in the simulation results in a nonmonotonic profile of the gravitational acceleration. The nonmonotonicity  leads to multiple solutions if the gravitational acceleration is used to predict the thermal properties of gas particles in simulated halos, while we find that the scalar force is not subject to the nonmonotonicity problem. We have also considered using the gravitational potential as one of the HPM variable, whose radial profiles of simulated halos are also monotonic. However, the gravitational potential suffers from the dynamic range being too narrow, and it has large-scale contributions that bias the local environment. After experimentation we found that using the gravitational potential is less optimal than the scalar force.

\subsection{HPM Table}\label{subsec:hpmtable}

\begin{figure*}[t]
\includegraphics[width=\textwidth]{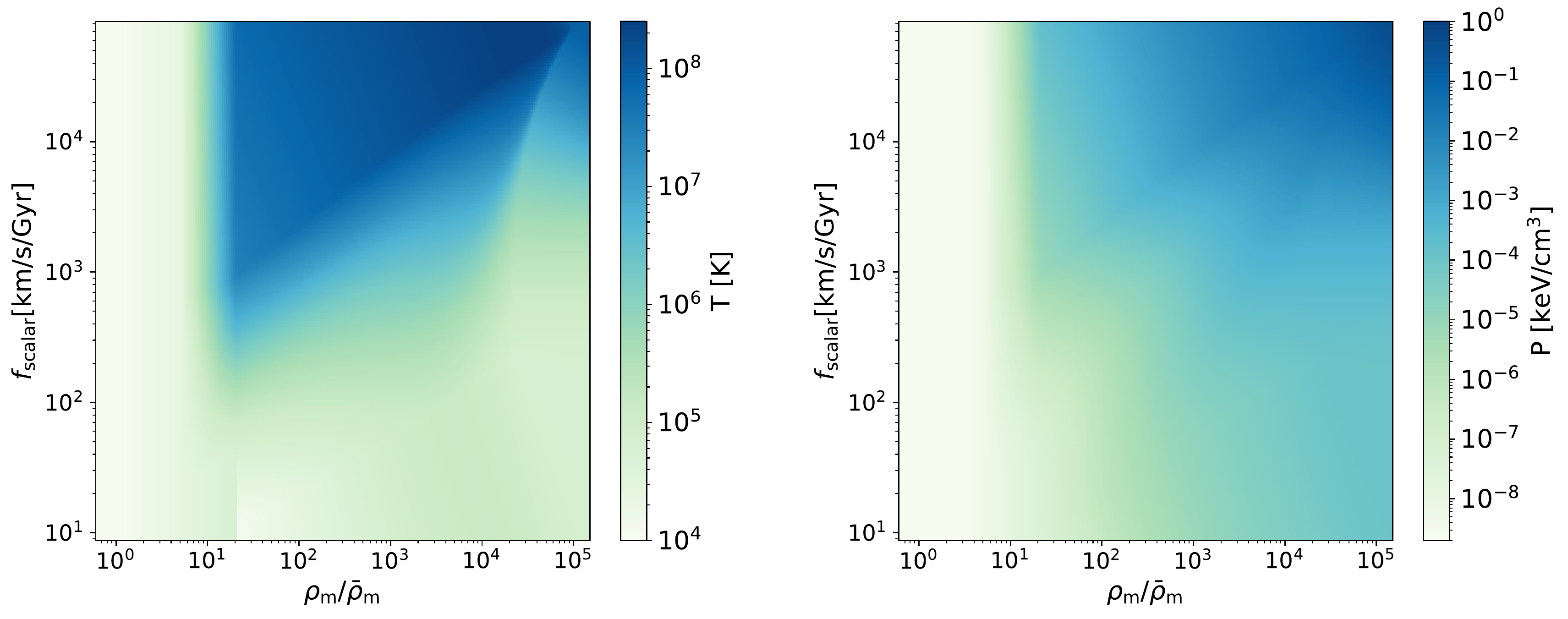}
\caption{Mapping relation from the HPM variables matter density, $\rho_{\rm m}$, and the scalar force, $f_\mathrm{scalar}$, discussed in Section~\ref{subsec:hpmvar}, to the gas temperature (left) and pressure (right). Both panels show that the HPM variables have a significant correlation with the target mapped quantities, though the mapping relation for the gas temperature has a more complicated pattern in the  top right corner due to the nonmonotonicity of the temperature profile in the core region of halos in the ICM model.}\label{fig:hpmtable}
\end{figure*}

With the HPM variables discussed in Section \ref{subsec:hpmvar}, matter density and scalar force, we aim to construct a mapping relation between the gas thermal properties and these designed halo variables. This mapping relation plays a crucial role in efficiently modelling the properties of gas in the HPM method for both the IGM and the ICM, and is referred to as the HPM table.

Construction of the HPM table in the low-density IGM regions where $\rho_{\rm m}/\bar{\rho}_{\rm m}\lesssim 10$ and in the high-density ICM region where $\rho_{\rm m}/\bar{\rho}_{\rm m}\gg 10$ in the simulation follows very different rules due to the dissimilar behavior of the gas in these two regimes. In the low-density IGM regime, the gas thermal properties can be very easily characterized through a power-law temperature-density relation mentioned in Section \ref{subsec:IGM} with just one of the HPM variables, the matter density $\rho_{\rm m}$. In the high density regime, the tight correlation between the gas density and the gas pressure breaks down, and we are no longer able to describe the density-temperature relation by a power-law.
However, with the help of the ICM model which introduces a mapping relation between the gas thermal properties and halo information, $M$ and $r$, we can construct a bridge between the HPM variables and the halo information and a mapping relation from the halo variables to the gas thermal properties.    

We construct a probabilistic relation between the gas thermal  properties, like temperature or pressure, and the designed HPM variables using a Bayesian analysis. Given the HPM variables for a gas particle $\rho_{\rm m}$ and $f_{\rm scalar}$, its average temperature or pressure can be estimated using a Monte Carlo approach:
\begin{equation}
\begin{split}
    &<X|\rho_{\rm m},f_{\rm scalar}> \\ &= \int X(M,r)p(M,r|\rho_{\rm m},f_{\rm scalar}) {\rm d}M{\rm d}r \\
    & =  \frac{\int X(M,r)p(\rho_{\rm m},f_{\rm scalar}|M,r)p(M,r){\rm d}M{\rm d}r}{\int p(\rho_{\rm m},f_{\rm scalar}|M,r)p(M,r){\rm d}M{\rm d}r} \\
    & \simeq \frac{\sum\limits_{M,r\sim p(M,r)}X(M,r)p(\rho_{\rm m},f_{\rm scalar}|M,r)}{\sum\limits_{M,r\sim p(M,r)}p(\rho_{\rm m},f_{\rm scalar}|M,r)} , 
\end{split}    
\end{equation}
and $X(M,r)$ is the gas pressure or temperature defined by the ICM model discussed in Section ~\ref{subsec:icmmodel}.

 In the continuous limit this mapping is deterministic, since $p(\rho,f|M,r)$ should be modelled as
 \begin{equation}
     p(\rho,f|M,r)=\delta(\rho-\rho_{\rm m}(M,r))\delta(f-f_{\rm scalar}(M,r)),
 \end{equation}
 where $\rho_{\rm m}(M,r)$ and $f_{\rm scalar}(M,r)$ are obtained from the ICM model with Eq.~\ref{eq:rho_m} and Eq.~\ref{eq:f_scalar}. In HYPER, we consider a discretized sampling method for our temperature estimation scheme, where we use a top-hat function to approximate $p(\rho_{\rm m},f_{\rm scalar}|M,r)$
 \begin{equation}\label{eq:halotable_tophat}
 \begin{split}
     &p(\rho_{\rm m},f_{\rm scalar}|M,r)= \Pi\left(\frac{x}{H}\right),\\
     &x=\sqrt{\left(\log\frac{\rho_{\rm m}}{\rho_{\rm m}(M,r)}\right)^2+\left(\log\frac{f_{\rm scalar}}{f_{\rm scalar}(M,r)}\right)^2} 
 \end{split}
 \end{equation}
where $H$ defines the width of the top-hat function. We note that  in the limit of a vanishing width of the top hat, we recover the aforementioned deterministic relation. 

For the prior distribution for $M,r$, we decompose the distribution as
\begin{equation}
\begin{split}
    p(M,r)&=p(M)p(r|M) \ ,
\end{split}
\end{equation}
where the prior for mass $p(M)$ and the conditional $p(r | M)$ are given as
\begin{equation}
\begin{split}
    p(M)&\propto M\frac{dn}{dM} \\
    p(r|M)&\propto \frac{r^2\rho_\mathrm{NFW}(M,r)}{M} \, .
    \end{split}
\end{equation}
Here $\frac{dn}{dM}$ is the halo mass function and $\rho_\mathrm{NFW}(M,r)$ is the halo density function defined by the NFW model.
Since we cannot directly sample from $p(M,r)$ without an on-the-fly halo finder, we apply the importance sampling, which is a general technique in statistics for estimating properties of a particular distribution while only having samples generated from a different distribution than the distribution of interest, for estimation of temperature or pressure, where $M,r$ are sampled from a uniform distribution on a logarithmic scale. We denote this covering distribution as $p_S(M,r)$. We can then re-weight the sampled points as
\begin{equation}\label{eq:T_estimator}
\begin{split}
        &<X|\rho_{\rm m},f_{\rm scalar}>\\
        &\simeq \frac{\sum\limits_{M,r\sim p_S(M,r)}X(M,r)p(\rho_{\rm m},f_{\rm scalar}|M,r)\frac{p(M,r)}{p_S(M,r)}}{\sum\limits_{M,r\sim p_S(M,r)}p(\rho_{\rm m},f_{\rm scalar}|M,r)\frac{p(M,r)}{p_S(M,r)}}.
\end{split}
\end{equation}

According to the analysis above, the procedure for building the HPM table consists of two steps:
\begin{itemize}
\item[1.] Building a two dimensional halo table, by imposing a grid whose columns and rows represent designed halo masses $M_i$ and radii $r_j$, where $M_i$ and $r_j$ are uniformly spaced within the mass and radius ranges on a logarithmic scale. This step simulates the process of sampling $M,r$ from a uniform distribution on a logarithmic scale. For each element in the ICM table, we can calculate the HPM variables, matter density $\rho_m$ and scalar force $f_{\rm scalar}$. Then we specify  $\hat{\rho}_{\rm m}^{(i,j)}=\rho_{\rm m}(M_i,r_j),\hat{f}_{\rm scalar}^{(i,j)}=f_{\rm scalar}(M_i,r_j)$.\\
\item[2.] Building the HPM table via HPM variables, a two dimensional table whose columns and rows represent the designed HPM variables $\rho_{\rm m}^{(k)}$ and $f_{\rm scalar}^{(l)}$, by traversing through the halo table we've built and looking into the designed distance between $(\rho_{\rm m}^{(k)},f_{\rm scalar}^{(l)})$ and $(\hat{\rho}_{\rm m}^{(i,j)},\hat{f}_{\rm scalar}^{(i,j)})$ defined in Eq.~\ref{eq:halotable_tophat}, we could relate the sampled points $(M_i,r_j)$  that would contribute to the estimator of the thermal quantities defined by Eq.~\ref{eq:T_estimator} corresponds to the mesh grid point $(\rho_{\rm m}^{(k)},f_{\rm scalar}^{(l)})$ in the HPM table. 
\end{itemize}

As mentioned above, we infer temperature and other thermal properties of the gas particles in simulation from a mapping relation between HPM variables and the thermal properties of gas particles we are interested in by
 \begin{equation}
     X = X(\rho_{\rm m},f_{\rm scalar}).
 \end{equation}
 For the intermediate region between the IGM and the ICM model we adopt a log-linear interpolation along the density axis of the  HPM table.
 
 In Figure~\ref{fig:hpmtable} we plot the HPM tables, which show the mapping relation between the HPM variables, the matter density $\rho_{\rm m}$ and the scalar force $f_{\rm scalar}$, and the gas temperature and pressure. In both panels of the plot, we see a vertical band along the axis of the scalar force at $\rho_{\rm m}/\bar{\rho}_{\rm m}\sim10$ that separates the HPM table into the IGM and the ICM  parts. We find the HPM table for the gas pressure generally follows a trend of larger pressure for particles with higher density and larger scalar force. However, the mapping relation for the gas temperature has a more complicated pattern in the top right region of the HPM table, where particles with high density and large scalar force are found. These particles are more likely to reside in the halo centers, where the  temperature profile is not necessarily monotonic.   

\clearpage
\subsection{Smoothing and Clumping Effects}

\label{subsec:sim_effect}
\begin{figure*}[hbt!]
\includegraphics[width=\textwidth]{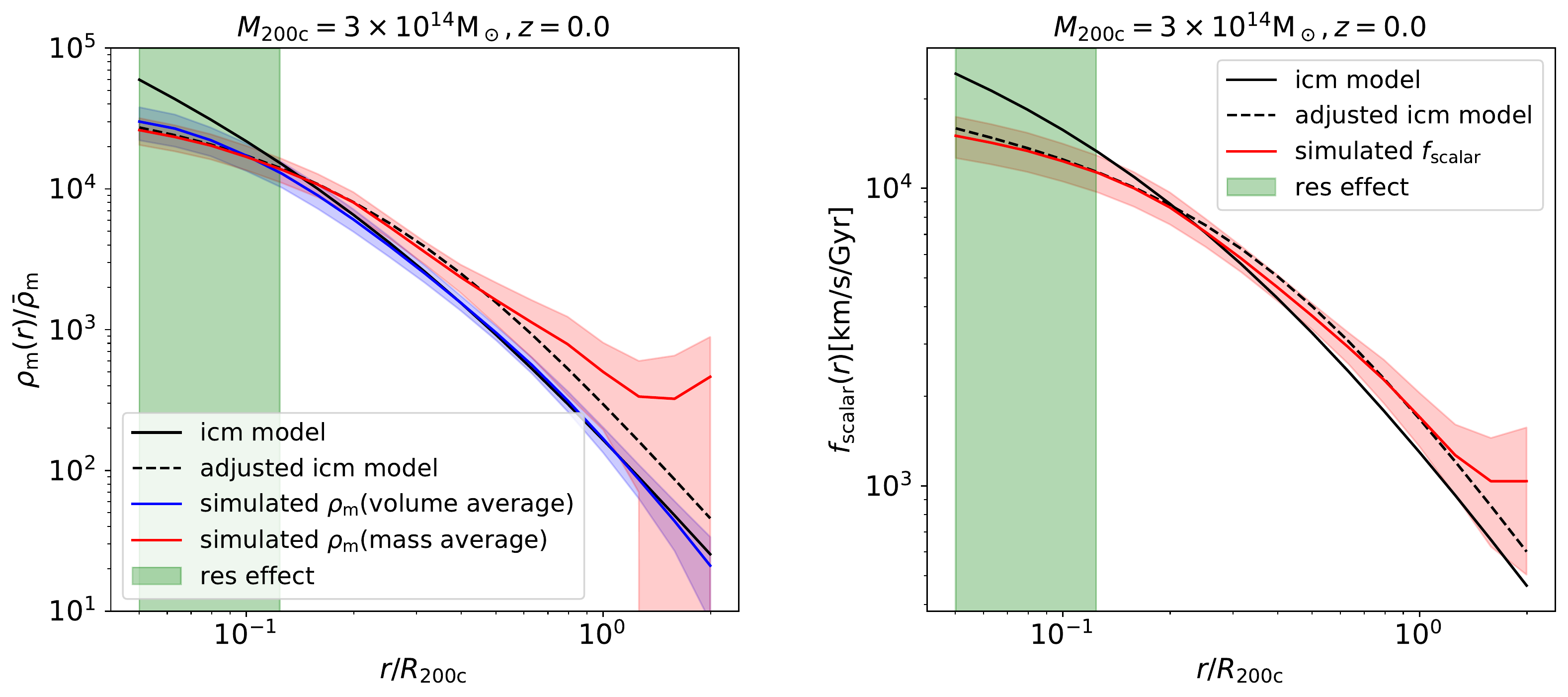}
\caption{Halo profiles of the HPM variables, matter density (left) and the scalar force (right), of halo mass bin centered at $3\times10^{14}{\rm M}_\odot$ at redshift $z=0.0$. Also shown is the resolution limit of the simulation (thin green band), where the simulated radial profiles of the HPM variables systematically deviate from the adopted ICM model. The volume weighted profiles of both the mass density and the scalar force of simulated halos (solid blue lines) are in good agreement with the NFW profile of the ICM model (solid black lines), except in the inner core, where they suffer from the limited resolution of the simulation. The mass weighted profiles (solid red lines) are greater than the ICM model at larger radii, and agree better with the adjusted ICM model described in the text (dashed black lines). The scatter of the volume weighted mass density profile (thin blue band) shows less dependence on the radius than the mass weighted one (thin red band).}\label{fig:hpmvars}
\end{figure*}

In the HPM simulation, due to the finite resolution of the mesh and the finite number of particles, the simulation results in the highest density regions like the inner cores of simulated halos suffer from a smoothing effect, which leads to underestimating output quantities of the simulation like the density and the scalar force. In addition, few of the simulated halos are completely relaxed (especially in the outskirts) or perfectly spherical, which violates the spherical symmetry assumption in the ICM model. We refer to the asymmetry and inhomogeneity in the distribution of particles in the outskirt region of simulated halos as the clumping effect. Both the smoothing effect and the clumping effect will cause the deviation of the simulated profiles for the HPM variables from the ICM model prediction. The smoothing effect will lead to an underestimation of simulated HPM variables calculated for the gas particles in the halo inner region, while the clumping effect in the outer region will lead to an overestimation.

In Figure \ref{fig:hpmvars} we show the comparison between the radial profiles of HPM variables, the matter density and the scalar force, and their ICM model predictions calculated for gas particles of simulated halos within the halo mass bin centered at $3\times10^{14}{\rm M}_\odot$ at redshift $z=0$. In both panels the simulation results tend to underestimate the HPM variables from the ICM model in the inner core region due to the smoothing effect of the finite numerical resolution of the simulation. Outside the core the volume weighted matter density profile is in good agreement with the NFW profile. We also find that the mass weighted profiles are significantly higher than the volume weighted ones, which indicates that the clumping effect in the outer regions of simulated halos is non-negligible.

\begin{figure}[hbt!]
\includegraphics[width=\hsize]{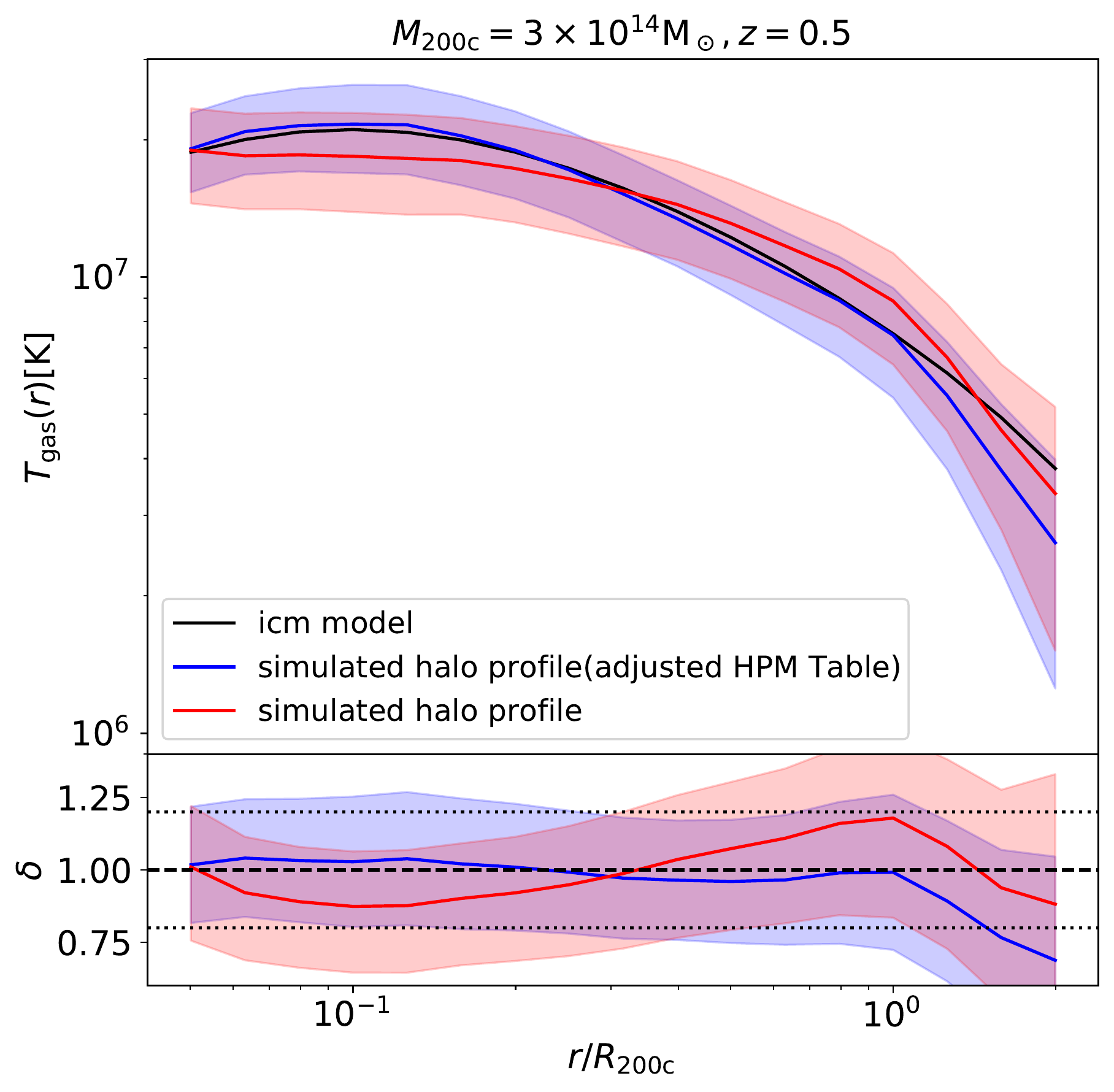}
\caption{{\bf Top:} Stacked temperature profiles of simulated halos (solid lines) and the scatter (thin bands) for the mass bin centered at $3\times10^{14}{\rm M}_\odot$ at redshift $z=0.5$. The profile measured in the simulation using the HPM table built on the original ICM model (red) underestimates the gas temperature in the inner region and overestimates the temperature in the outskirts up to 20\%. The temperature profile measured in the simulation that uses the HPM table adjusted for smoothing and clumping effect (blue) agrees better with the temperature profile of the ICM model (black). {\bf Bottom:} Relative difference of simulated profiles with the ICM model and a $\pm20\%$ region (dotted black lines)}\label{fig:adjusted_hpmtable}
\end{figure}

Notice that since we infer the gas pressure or temperature from the HPM variables calculated for each particle, the offset between the simulation results and the ICM model would introduce bias to the inference of gas temperature or pressure with the HPM table and will result in incorrect dynamics in simulation. We must treat the smoothing and clumping effects carefully when using the HPM variables calculated for each gas particle to model their thermal properties in the simulation. In Figure \ref{fig:adjusted_hpmtable} we plot the temperature profiles of simulated halos within the mass bin centered at $M_\mathrm{200c}=3\times10^{14}{\rm M}_\odot$ at redshift $z=0.5$, where the temperature of gas particles are interpolated from the HPM table built with the original ICM model. Ignoring the smoothing and clumping effects on HPM variables results in underestimating the temperature of gas particles in the center of the halo and an overestimating it in the outer region, and the deviation from the ICM model prediction on the gas temperature could be up to 20\%.

The reason for this mismatch is that the smoothing effect in the inner region of simulated halos suppresses magnitude of the HPM variables to the values below the ICM model prediction. According to the HPM table we show in Figure~\ref{fig:hpmtable}, the HPM variables have a significantly positive correlation with the temperature. Thus, underestimating the HPM variables means we are also very likely to underestimate the inferred temperature simultaneously. Then gas particles that reside in the halo center are more likely to be assigned a temperature value that is biased low. In the outer region of simulated halos the clumping effect leads to HPM variables for a substantial number of particles being higher than the ICM prediction. Due to the same reason that temperature and HPM variables are positively correlated in the HPM table, overestimating of the HPM variables leads to the temperature of gas being too high in the outer region.

To mitigate this inconsistency in the temperature inference caused by the limited resolution and break down of spherical symmetry, we need to take the smoothing and clumping effects into account while building the halo table. We first run a dark matter only simulation and fit a calibration function $C(r,\zeta)$ to the ratio of simulated profiles of HPM variables to their ICM model predictions, which enables us to capture the differences between the simulation results and the original ICM model. We have discussed in Section~\ref{subsec:hpmvar} that the HPM variables are only mildly affected by the gas distribution, so the profiles of HPM variables derived from the dark matter only simulation can properly emulate their values in an HPM simulation. We verified that the offset between simulated profiles of HPM variables and their ICM model predictions measured in the dark matter only simulation and HPM simulation were indeed very similar. 

The calibration function measured from a dark matter only simulation can effectively correct the gas temperature or pressure inference in HYPER. We adjust the calculation with the fitted calibration function when deriving the HPM variables in the halo table as 
\begin{equation}
\begin{split}
   \hat{\rho}_{\rm m}^{(i,j)}&=\rho_{\rm m}(M_i,r_j)*C_{\rho_{\rm m}}(r_j,\zeta),\\
   \hat{f}_{\rm scalar}^{(i,j)}&=f_{\rm scalar}(M_i,r_j)*C_{f_{\rm scalar}}(r_j,\zeta), 
\end{split}
\end{equation}
 where $\zeta$ refers to any information needed to specify numerical resolution. The HPM table built with the adjusted halo table more accurately relates the HPM variables of simulated particles to their thermal properties and better approximates the hydrodynamics of gas particles through the ICM model. The profiles of HPM variables adjusted for the calibration functions are also shown in Figure \ref{fig:hpmvars}. More details about the calibration functions are presented in Appendix~\ref{subsec:halovar_adjust}.

After adjusting the HPM table for the effects mentioned above, we show in Figure \ref{fig:adjusted_hpmtable} that the temperature inferred from the adjusted HPM table is in good agreement with the ICM prediction. We observe a downturn in the radial temperature profile beyond $R_\mathrm{200c}$, which is due to the existence of a substantial amount of low-temperature IGM gas at the outskirt region of the massive halos.

\subsection{Artificial Viscosity and Pressure Filtering}\label{subsec:filter_vsc}
Shocks are a generic feature of gas flows. When solving the fluid equations in a particle-based simulation like SPH, the entropy generation on shocks is captured by an artificial viscosity term. HYPER is a Lagrangian particle method for the dynamical evolution of gas similar to SPH, and hence needs an artificial velocity term in the gas momentum equation:
\begin{equation}\label{eq:gas_motion_viscosity}
    \frac{{\rm d}\vec{v}_\mathrm{gas}}{{\rm d}\tau}=-\nabla\Phi-\frac{\nabla P}{\rho_\mathrm{gas}}-\vec{a}^\mathrm{visc}.
\end{equation}
Though we don't need to resolve the conversion between kinetic energy and thermal energy in HYPER as the thermal properties of gas are directly inferred from the HPM table built on the ICM model, we still need to include the artificial viscosity term to prevent particle interpenetration in shocks. The artificial viscosity in HYPER removes the part of the kinetic energy that should be converted into heat. Otherwise, we won't be able to accurately resolve the motion and distribution of the gas in our simulation (Appendix \ref{subsec:filter_vsc_tune}). Many different forms have been suggested for the artificial viscosity, with the Von Neuman-Richtmyer artificial viscosity being the simplest one:
\begin{equation}
    \vec{a}^\mathrm{visc}=\frac{\nabla Q}{\rho_\mathrm{gas}}
\end{equation}
with
\begin{equation}\label{eq:artificial_viscosity_Q}
    Q=\begin{cases}
    \alpha h\rho_\mathrm{gas} c|\nabla\cdot\vec{v}|+\beta h^2\rho_\mathrm{gas}|\nabla\cdot\vec{v}|^2 & \nabla\cdot\vec{v}<0,\\
    0 & \nabla\cdot\vec{v}\geq0,
    \end{cases}
\end{equation}
where $h$ is the particle smoothing length, proportional to the local mean inter-particle separation,  $h\propto \rho_\mathrm{gas}^{-1/3}$; $c$ is the speed of sound of gas. In this work, $\alpha\sim 0.1, \beta\sim 0.05$ are found to be the best fit values when calibrating the gas radial profiles and scaling relations of different integrated quantities of the simulated halos. The viscosity term we adopt has a similar form as the gas pressure force term $\nabla P/\rho_\mathrm{gas}$, with $Q$ acting as an excess pressure assigned to gas particles. To integrate the artificial viscosity in HYPER, we only need to solve the dynamical equation for gas particles \ref{eq:gas_motion} with the modified pressure $\nabla (P+Q)/\rho_\mathrm{gas}$. 

In our HPM simulation, in addition to adopting the artificial viscosity for gas particles, we also apply a force filter on the pressure force $\nabla P/\rho_\mathrm{gas}$. This approach aims to sustain a proper hydrostatic equilibrium in the central core region of simulated halos, since the interpolated temperature of gas particles in the center of the halos suffers is not affected as much by the smoothing effect of finite resolution as the gas density, and that in turn leads to less smoothing of the gas pressure compared to the gravitational force. Filtering on the small-scale structure of the pressure field will impose a same degree of smoothing effect on the hydro force as the gravitational force in the center region of the simulated halos. Filtering also suppresses the numerical noise introduced by the interpolation of gas temperature from a model-based HPM table. We filter the gas pressure field in the Fourier space by
\begin{equation}
    \tilde{P}^f(\vec{k})=\tilde{P}(\vec{k})f(\vec{k})
\end{equation}
and the filtered pressure force is given by replacing the gas pressure $P$ in $\nabla P/\rho_{\rm gas}$ with the filtered pressure $P^f$. For the specific functional form for the filtering we adopt the Weibull function
\begin{equation}\label{eq:p_filter}
    f(\vec{k})=A_{\rm L}^f-(A_{\rm L}^f-A_{\rm H}^f)\exp[-(A_{\rm S}^fk)],
\end{equation}
where $A_{\rm H}^f = 1.0$ for the high frequency filter. Parameters $A_{\rm L}^f$ and $A_{\rm S}^f$ are tuned to match the simulated results with the ICM model predictions: $A_{\rm L}^f=0.5, A_{\rm S}^f=10.0$.

A combination of the artificial viscosity and filtering on gas pressure helps to prevent too much gas from being ejected of the halos in our simulation and to ensure we get a reasonable gas fraction for the simulated halos. Tuning the small scale filter and artificial viscosity are presented in Appendix~\ref{subsec:filter_vsc_tune}.

\section{Results} \label{sec:results}

In this section we compare the results from a HYPER simulation to the predictions of the ICM model used to implement the HPM mapping relation for gas thermal properties. Good consistency in the properties of the ICM such as the radial profiles, scaling relation of integrated halo properties, and measurements of the tSZ effect implies one can systematically control the ICM physics in HYPER simulation by varying the ICM model while constructing the HPM mapping relation.

\subsection{Halo Radial Profiles}\label{subsec:profiles}
\begin{figure*}[hbt!]
\begin{center}
\includegraphics[width=\textwidth]{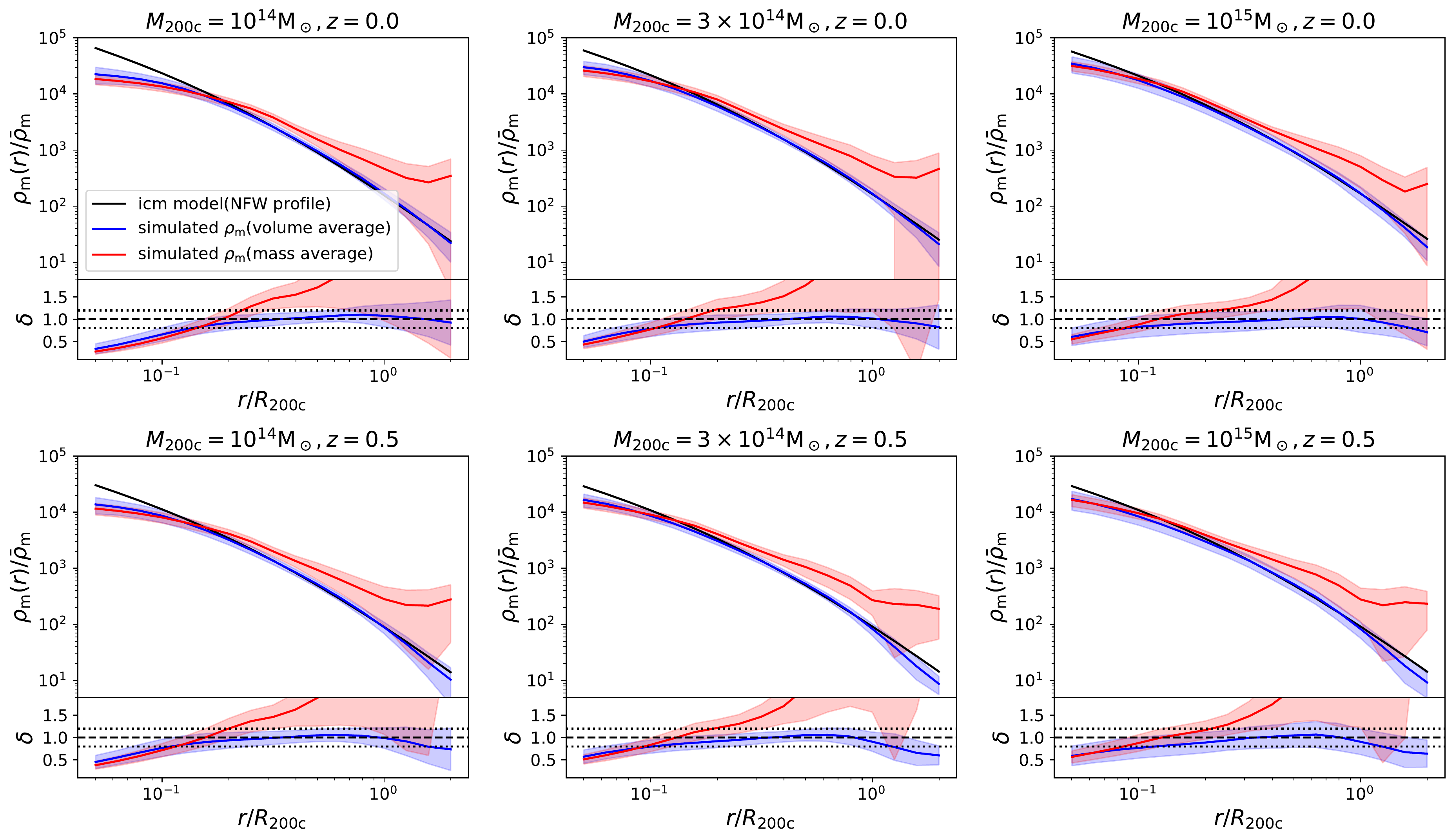}
\caption{Comparison between the simulation results and the ICM model (NFW profile) for the radial profiles of matter density for halos in mass bins centered at $10^{14}{\rm{M}}_\odot$ (1st column), $3\times10^{14}{\rm{M}}_\odot$ (2nd column), and $10^{15}{\rm{M}}_\odot$ (3rd column) at redshift $z=0$ (1st row) and $z=0.5$ (2nd row). In each plot the top panel shows that the simulated volume weighted matter density profile (blue line) agrees well with the NFW profile (black line). In contrast, the mass weighted matter density profile (red line) overestimates the matter density in the outskirt region compared to the standard ICM model due to the clumping effect. Also shown are the scatter in the simulated matter density profiles (thin blue/red bands); the bottom panel shows the ratio between the simulation results and the ICM model and a $\pm20\%$ region (dotted black lines).}\label{fig:rhoM_profiles}
\end{center}
\end{figure*}

\begin{figure*}[hbt!]
\begin{center}
\includegraphics[width=\textwidth]{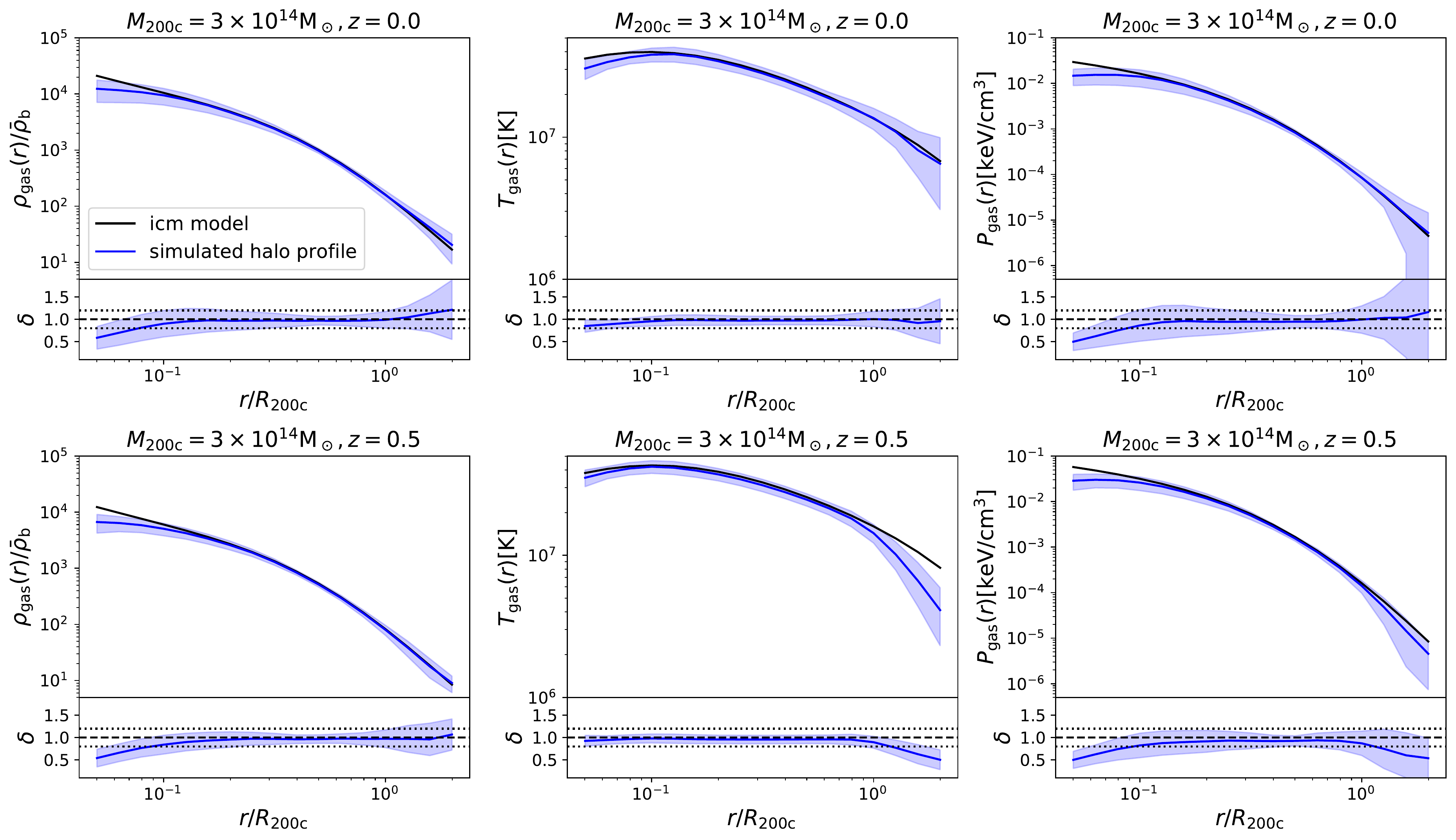}
\caption{Comparison between the simulation results and the ICM model prediction for the volume weighted radial profiles of gas density (1st column) and pressure (3rd column) and mass-weighted temperature (2nd column) for halos in mass bins centered at $3\times10^{14}{\rm{M}}_\odot$ at redshift $z=0$ (1st row) and $z=0.5$ (2nd row). For each plot, the top panel shows the simulated halo profile and its scatter (blue line and band), which is found to be in good agreement with the profile of the ICM model (black line); the bottom panel shows the ratio between the simulation results and the ICM model and a $\pm20\%$ region (dotted black lines). }\label{fig:halo_profiles}
\end{center}
\end{figure*}
We first compare the profiles of matter density and gas thermal properties of simulated halos to the profiles derived with the ICM model. In Figure~\ref{fig:rhoM_profiles} we show the radial profiles of matter density  and their scatter for the simulated halos whose masses fall into three mass bin centered at $10^{14}{\rm{M}}_\odot$, $3\times10^{14}{\rm{M}}_\odot$, and $10^{15}{\rm{M}}_\odot$ at redshift $z=0$ and $z=0.5$. We find that the volume-weighted density profiles agree with the NFW model well except in the inner region, where they suffer from the smoothing effect due to the limited resolution of the simulation. The clumping effect in the mass-weighted density profiles mentioned in Section~\ref{subsec:sim_effect} leads to an overestimate of the matter density at the outskirts of halos.

In Figure~\ref{fig:halo_profiles} we plot the simulated profiles of gas density, temperature, and pressure and their scatter for the halos whose masses fall into the mass bin centered at $3\times10^{14}{\rm{M}}_\odot$ at redshift $z=0$ and $z=0.5$. We also plot the prediction for the gas profiles in the ICM model used for implementing the hydro part of the simulation for comparison, and we also show the ratio of simulated profiles to the ICM model predictions in the bottom panels. We find that the radial profiles of the gas density, temperature, and pressure for simulated halos are in about 5\% agreement with the ICM model from 0.1$R_\mathrm{200c}$ to $R_\mathrm{200c}$ and remain within 20$\%$ even in the inner core and in halo outskirts up to about 1.5$R_\mathrm{200c}$. 

In the inner core we observe the smoothing effect due to the limited resolution of the simulation in the gas density and pressure profiles. An analogous smoothing effect in the simulated gas temperature profiles in the halo centers is much less significant because we already account for the numerical resolution effect when constructing the HPM table.

\subsection{Integrated Halo Quantities}\label{subsec:halo quantities}

\begin{figure*}[hbt!]
\begin{center}
\includegraphics[width=1.0\textwidth]{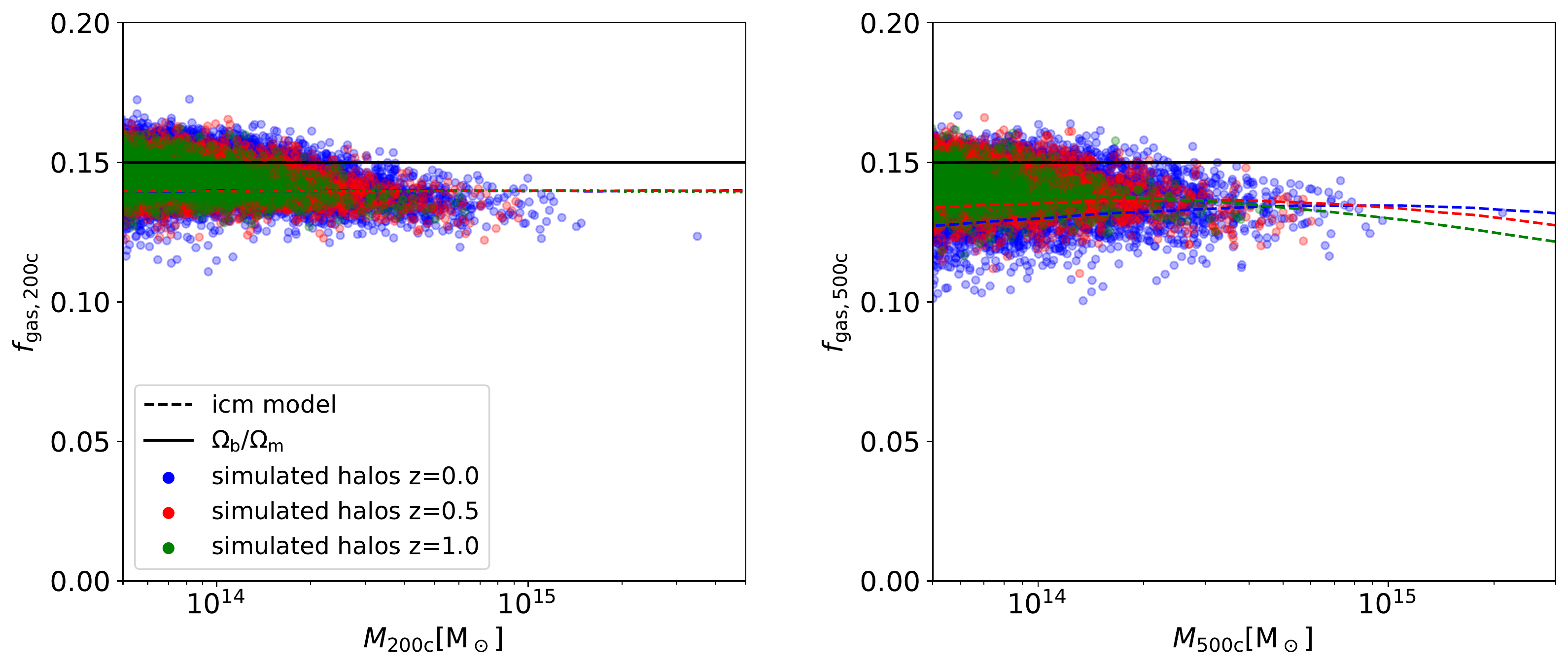}
\caption{Gas fraction of simulated halos (dots) compared to the ICM model prediction (dashed lines) at different redshift $z=0$ (blue), $z=0.5$ (red), and $z=1.0$ (green) within a spherical region with the mean overdensity 200 (left) and 500 (right) times of critical density. The halo gas fractions in the simulation are substantially lower than the universal baryonic fraction $\Omega_{\rm b}/\Omega_{\rm m}=0.15$ for the adopted cosmology (solid black line).}\label{fig:f_gas}
\end{center}
\end{figure*}

X-ray observables, such as luminosity, temperature,
mass of the ICM, and SZ flux of galaxy clusters have been proposed and used as proxies for the total cluster mass \citep[e.g.][]{2005RvMP...77..207V}. Calibrating relations between cluster mass and these observables is important for exploiting the full statistical power of the cluster surveys. In this section we show further comparison between the simulation results and the ICM model predictions by investigating the integrated quantities of identified halos and exploring the scaling relation between the SZ effect signal, X-ray observables and halo masses in the simulation.

In Figure \ref{fig:f_gas}, we plot the gas fractions of simulated halos enclosed within spheres with the averaged density of 200 and 500 times the critical density respectively at different redshifts $z=0, 0.5, 1.0$. For comparison, we also plot the gas fraction derived from the ICM model and the universal baryonic fraction $f_{\rm b}=\Omega_{\rm b}/\Omega_{\rm m}$ where $\Omega_{\rm b}$ and $\Omega_{\rm m}$ are the cosmological parameters set for the simulation. As shown in the plots, after implementation of gas pressure in solving motion of baryonic components in the simulation, the gas fraction of simulated halos are substantially lower than the fraction baryon mass takes of the mass of all the matters in the universe, which means a considerable portion of gas gets propelled out of the gravitational potential well of collapsed halos due to existence of gas pressure. Furthermore, we find simulation results of gas fraction match the prediction derived from analytical models of gas pressure profiles by assuming hydrostatic equilibrium reasonably well. Scatter in the results of simulated halos is not negligible, as the dynamical state of different halos can vary significantly in a real simulation. We observe slightly more scatter inside the radius encompassing the overdensity of 500 times the critical density, since the inner region is more sensitive to the dynamic state of the halos. The failure of simulation to resolve the inner cores of halos may also contribute to the overall scatter.           

\begin{figure*}[hbt!]
\begin{center}
\includegraphics[width=1.0\textwidth]{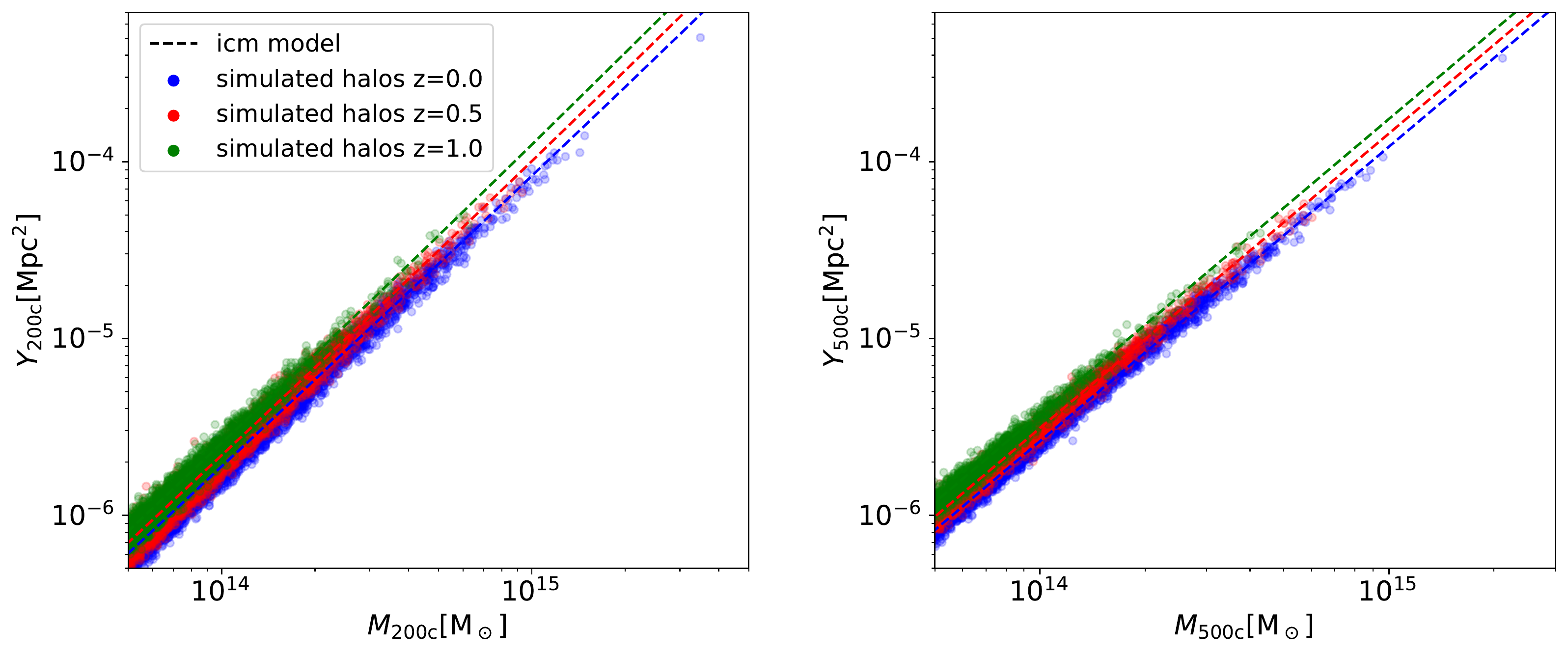}
\caption{Integrated Compton-$Y$ parameter within a spherical region with the mean overdensity of 200 (left) and 500 (right) times the critical density for simulated halos (dots) compare with the ICM model prediction (dashed lines) at different redshifts $z=0$ (blue), $z=0.5$ (red) and $z=1$ (green).}\label{fig:Compton_obs}
\end{center}
\end{figure*}

In Figure \ref{fig:Compton_obs}, we plot the Compton $Y$-parameter integrated over a spherical volume,
\begin{equation}
    Y_\mathrm{R}=\frac{\sigma_\mathrm{T}}{m_\mathrm{e}c^2}\int_{0}^{R}4\pi P_\mathrm{e}(r)r^2{\rm d}r,
\end{equation}
for the simulated halos and compare it with the $Y-M$ relation derived from the gas pressure model for two different mass definitions at different redshifts. The Compton parameter $Y_{\rm{200c}}$ and $Y_{\rm{500c}}$ integrated within a spherical region with the mean overdensity of 200 and 500 times the critical density of simulated halos is consistent with the $Y-M$ relation derived from the gas pressure model, which is not surprising since Figure \ref{fig:halo_profiles} shows that the simulated gas pressure profiles agree  well with the analytical pressure model. Scatter in the $Y-M$ scaling relation in the simulation is smaller than the scatter in gas fractions and other X-ray observables; apparently the Compton $Y$ signal of simulated halos is less affected by the poorly resolved inner core region. Our results agree with the conclusion in other studies of the SZ effect \citep[e.g.][]{2002MNRAS.336.1256K} that the Compton $Y$ signal is less sensitive to gas physics in the core of clusters.

\begin{figure*}[hbt!]
\begin{center}
\includegraphics[width=\textwidth]{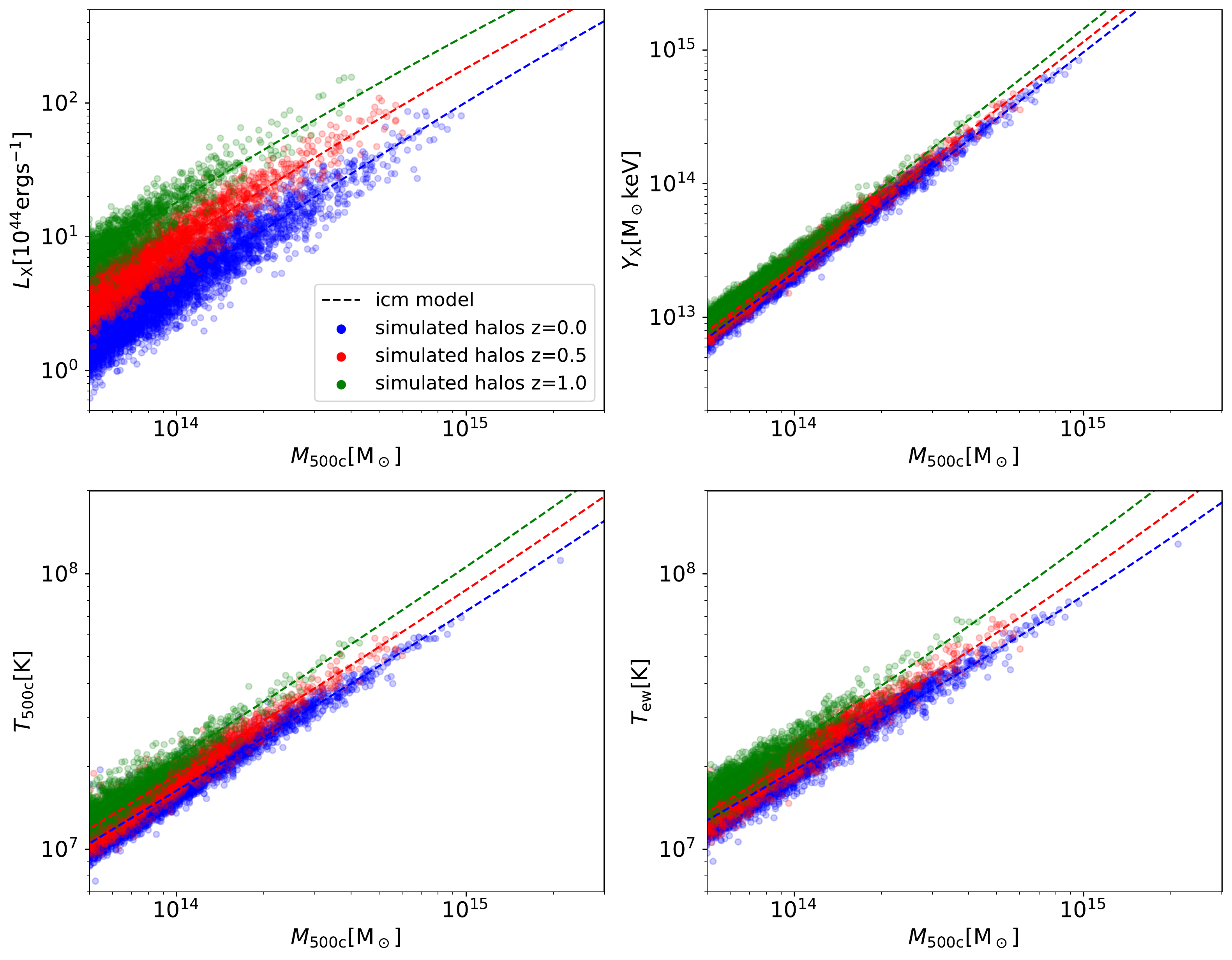}
\caption{Integrated X-ray quantities: the bolometric X-ray luminosity (top left), the spherical Compton-like $Y_\mathrm{x}$ parameter (top right), the mass-weighted average temperature (bottom left), and the emission-weighted temperature (bottom right) of the simulated halos (dots) compared with the ICM model prediction (dashed lines) at different redshifts $z=0$ (blue), $z=0.5$ (red) and $z=1$ (green).}\label{fig:X-ray_obs}
\end{center}
\end{figure*}
In Figure \ref{fig:X-ray_obs} we plot the bolometric X-ray luminosity $L_\mathrm{X}$, 
\begin{equation}
L_\mathrm{X}=\int\Lambda(T)n_\mathrm{gas}^2{\rm d}V,
\end{equation}
where $\Lambda(T)\propto\sqrt{T}$ is the cooling function assuming that Bremsstrahlung emission dominates, and $n_\mathrm{gas}$ is the gas density. We also plot the emission weighted temperature $T_\mathrm{ew}$ 
\begin{equation}
T_\mathrm{ew}=\frac{\int\Lambda(T)n_\mathrm{gas}^2T{\rm d}V}{\int\Lambda(T)n_\mathrm{gas}^2{\rm d}V}.
\end{equation}
Another common X-ray observable is the $Y_{\rm X}$ parameter \citep{2006ApJ...650..128K}, proportional to the gas thermal energy as defined by the product of the gas mass and the spectroscopic X-ray temperature,
\begin{equation}
    Y_\mathrm{X}=M_\mathrm{gas}T_{\rm{X}},
\end{equation}
where $M_\mathrm{gas}$ is the gas mass within the spherical overdensity region of radius $R_\mathrm{500c}$ and $T_\mathrm{X}$ is the spectroscopic X-ray temperature. Finally, we also show in Figure \ref{fig:X-ray_obs} the characteristic temperature $T_\mathrm{500c}$,
\begin{equation}
    T_\mathrm{500c}=\frac{\mu_\mathrm{e}m_\mathrm{e}m_\mathrm{p}c^2}{k_\mathrm{B}\sigma_\mathrm{T}}\frac{Y_\mathrm{500c}}{M_\mathrm{gas}}.
\end{equation}

Scaling relation of the integrated X-ray quantities and halo mass in the simulation results are in good agreement with the ICM gas model predictions from redshift $z=0$ to $z=1$. Notice that all these X-ray observables are exclusively defined by the density and the temperature of the gas in halos. Since we have shown in Figure \ref{fig:halo_profiles} that the radial profiles of both gas density and temperature are in good agreement with the analytical model, this explains the consistency between the integrated X-ray quantities of the simulated halos and the ICM model prediction. Simulation results for the $Y_\mathrm{X}-M$ relation have been shown to have less scatter than the $L_\mathrm{X}-M$ relation, which may be due to the anti-correlation between the deviation of $M_\mathrm{gas}$ and $T_\mathrm{X}$ from the scaling relation prediction, which has been found in X-ray observations \citep{2006ApJ...650..128K}.

\begin{figure}[hbt!]
\begin{center}
\includegraphics[width=\hsize]{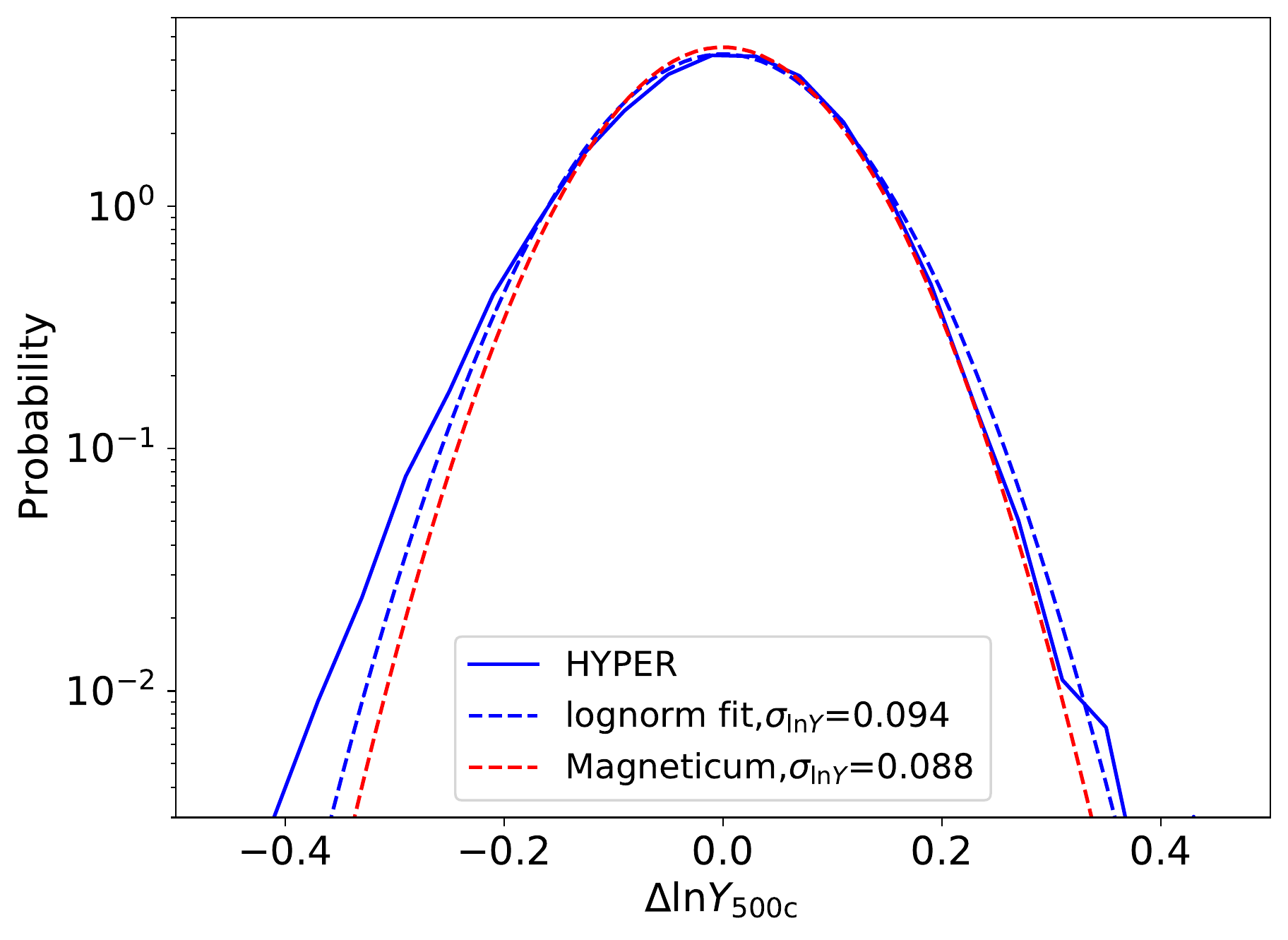}
\caption{Probability distribution of the scatter about the $Y_{\rm{500c}}$-$M_{\rm{500c}}$ relation for the HYPER simulation output at redshift $z=0.0$ (solid blue line), and its best lognormal fit (dashed blue line). The lognormal fit to the Magneticum simulation output at redshift $z=0.0$ (dashed red line) is also shown for comparison and agrees well with the HYPER results.}\label{fig:scaling_relation_scatter}
\end{center}
\end{figure}

We also look into the scatter about the scaling relation in HYPER simulation, as it could reflect the information on important dynamical effects. \citet{2017MNRAS.469.3069G} studies the scatter about $Y-M$ relation in the Magneticum simulation and finds that the scatter $\sigma_{{\rm{ln}}Y}$ about the mass-observable relations at overdensity $\rm{500c}$ is consistent with the lognormal distribution. We also calculate the distribution of scatter about the $Y_{\rm{500c}}$-$M_{\rm{500c}}$ relation in HYPER and fit it to a lognormal distribution. We find that the rms scatter $\sigma_{{\rm{ln}}Y}=0.094\pm0.002$ is in good agreement with the value $\sigma_{{\rm{ln}}Y}=0.088\pm0.006$ from the Magneticum simulation. We show the probability distributions of scatter about the $Y_{\rm{500c}}$-$M_{\rm{500c}}$ for HYPER and its best lognormal fit, as well as the best-fit lognormal from the Magneticum simulation in Figure~\ref{fig:scaling_relation_scatter}. We also measure the scatter about the X-ray observable-mass relation by calculating the RMS dispersion
\begin{equation}
    \sigma_{\log_{10}Y}=\sqrt{\frac{1}{N}\sum_{i=1}^N\left[\log_{10}(Y_i)-\log_{10}(Y_{\rm model})\right]^2}
\end{equation}
following \citet{2017MNRAS.465..213B}, where $Y_i$ is the X-ray observables as $i$ runs over all simulated halos and $Y_{\rm model}$ is the corresponding ICM model prediction. We present the comparison between our measurement for the scatter about $L_{\rm X}-M_{\rm 500c}$, $Y_{\rm X}-M_{\rm 500c}$, and $T_{\rm{500c}}-M_{\rm{500c}}$ and the MACSIS simulation results in Table~\ref{tab:X-ray scatter}.

\begin{table}
\begin{center}
\begin{tabular}
{ p{1.2cm}||p{1.7cm}|p{1.7cm}|p{2.0cm} }
 \hline
 $\sigma_{{\rm{\log_{10}}}Y}$ &$L_{\rm{X}}$-$M_{\rm{500c}}$ &$Y_{\rm{X}}$-$M_{\rm{500c}}$ &$T_{\rm{500c}}$-$M_{\rm{500c}}$ \\
 \hline
 HYPER &0.153 &0.083 &0.038 \\
 \hline
  MACSIS &0.15 $\pm$0.02 &0.12 $\pm$0.01 &0.048 $\pm$0.003 \\
 \hline
\end{tabular}\caption{Scatter about the X-ray observable - mass relation for HYPER and MACSIS at redshift $z=0$.}\label{tab:X-ray scatter}
\end{center}
\end{table}
We find the scatter about the scaling relation in HYPER is in general consistent with the state-of-art full hydro simulations, suggesting that HYPER captures the important dynamical effects modeled in full simulations. However, to further verify this point, we need to adopt the ICM model drawn from the full hydro simulations to construct the HPM table for inferring the gas thermal properties. Studies show that the ICM physics could also affect the scatter about the scaling relation \citep[e.g.][]{2012ApJ...758...74B}. The excellent agreement on the scatter about the $Y_{\rm{500c}}-M_{\rm{500c}}$ relation between HYPER and Magneticum may result from the ICM gas pressure model we adopt being similar to that found in the Magneticum simulation, which implies HYPER adopting current ICM model may properly emulates the ICM physics in Magneticum simulation and the good match on the scatter is a result of a more fair comparison. Hence, more detailed comparison for scatter about the observable-mass relation of HYPER and full hydrodynamic simulations is required in the future study.

\subsection{Thermal SZ Angular Power Spectrum}\label{subsec:tSZ_power}

The tSZ power spectrum $C_l$ is a powerful probe of cosmology and can provide promising constraints on cosmological parameters, in particular $\sigma_8$, since $C_l\propto \sigma_\text{8}^{7-9}$ \citep[e.g.][]{2002MNRAS.336.1256K, 0004-637X-725-2-1452,2011ApJ...727...94T}. Because the cluster signal dominates tSZ anisotropies, we can model the analytical prediction of the tSZ power spectrum using the standard halo model \citep[e.g.][]{1988MNRAS.233..637C, 1538-4357-526-1-L1}. The tSZ angular power spectrum at a multipole moment $l$ for the one-halo term is given by 
\begin{equation}\label{eq:tSZ}
C_l^\mathrm{tSZ}=f^2(\nu)\int_z\frac{\mathrm dV}{\mathrm dz}\int_M\frac{\mathrm dn(M,z)}{\mathrm dM}|\tilde{Y}_l(M,z)|^2{{\rm d}M{\rm d}z},
\end{equation}
where $f(\nu)=x_\nu\coth(x_\nu/2)-4$ with $x_\nu=h\nu/(k_\text{B}T_\text{CMB})$ is the spectral shape of the tSZ signal. Integration over redshift and mass are carried out from $z=0$ to $z=5$ and from $M=10^{10}{\rm{M}}_{\odot}$ to $M=10^{16}{\rm{M}}_{\odot}$ respectively. For the differential halo mass function $\mathrm dn(M,z)/\mathrm dM$ we adopt the fitting function from \citet{2008ApJ...688..709T} based on N-body simulations. 

Following \citet{2002MNRAS.336.1256K}, the 2D Fourier transform of the projected Compton $Y$-parameter, $\tilde{Y}_l(M,z)$ is given in the Limber approximation \citep{1953ApJ...117..134L} as
\begin{equation}\label{eq:29}
\tilde{Y}_l(M,z)=\frac{4\pi r_\text{s}}{l_\text{s}^2}\frac{\sigma_\text{T}}{m_\text{e}c^2}\int x^2P_\text{e}(x)\frac{\sin(lx/l_\text{s})}{lx/l_\text{s}}\mathrm dx,
\end{equation}
where $x=r/r_\text{s}$ is a scaled dimensionless radius, $r_\text{s}$ is characteristic radius for a NFW profile defined as $R_\text{500c}/c_\text{500c}$, and we use the average halo concentration $c_\text{500c}$ calibrated as a function of the cluster mass and redshift from \citet{2015ApJ...799..108D}. The corresponding angular wave number $l_\text{s}=d_\text{A}/r_\text{s}$, where $d_\text{A}(z)$ is the proper angular-diameter distance at redshift $z$. Here $P_\text{e}(r)=P_\text{th}(r)\mu/\mu_\text{e}$ is the electron pressure, and $\mu$ and $\mu_\text{e}$ are the mean mass per gas particle and per electron respectively, and $P_\text{th}(r)$ is the gas thermal pressure profile of the ICM model we discuss in Section~\ref{sec:models}. The integral is carried out within a spherical region with radius $R\sim 4R_\mathrm{500c}$.

In the Limber approximation we can also relate the thermal SZ angular power spectrum to the 3D thermal pressure power spectrum by
\begin{equation}\label{eq:tsz_p}
    C_l^\mathrm{tSZ}=\frac{16f^2(\nu)\pi^2}{(2l+1)^3}\int_{0}^{z_\mathrm{max}}\Delta_\text{tSZ}(k,z)|_{k=l/\chi}\chi(z)d\chi(z),
\end{equation}
where $z_\mathrm{max}=5$, $\chi(z)$ is the comoving distance to redshift $z$, and
\begin{equation}
    \Delta_\text{tSZ}(k,z)=\left[\frac{\langle P_{\rm e}(z)\rangle\sigma_\text{T}}{(1+z)m_\text{e}c^2}\right]^2\frac{k^3}{2\pi^2}P_p(k,z).
\end{equation}
Here $P_p(k,z)=\langle\delta_P(\vec{k},z)\delta_P^*(\vec{k},z)\rangle$ is the power spectrum of the Fourier transform $\delta_P(\vec{k},z)$ of the fractional thermal pressure fluctuations $\delta_P(\vec{x},z)\equiv P_{\rm e}(\vec{x},z)/\langle P_{\rm e}(\vec{x},z)\rangle-1$. We approximate the tSZ power spectrum as a sum over the finite number of simulation outputs,
\begin{equation}\label{eq:tsz_p_tomo}
    C_l^\mathrm{tSZ}=\frac{16f^2(\nu)\pi^2}{(2l+1)^3}\sum_{i}\Delta_\text{tSZ}(l/\chi_i,z_i)\chi_i\Delta\chi_i,
\end{equation}
and we use the fast Fourier transform to calculate the power spectrum of the gas pressure field drawn from the simulation snapshots at redshift $z_i$ to directly estimate the thermal SZ angular power spectrum from our simulation output.

\begin{figure}[th!]
\includegraphics[width=\hsize]{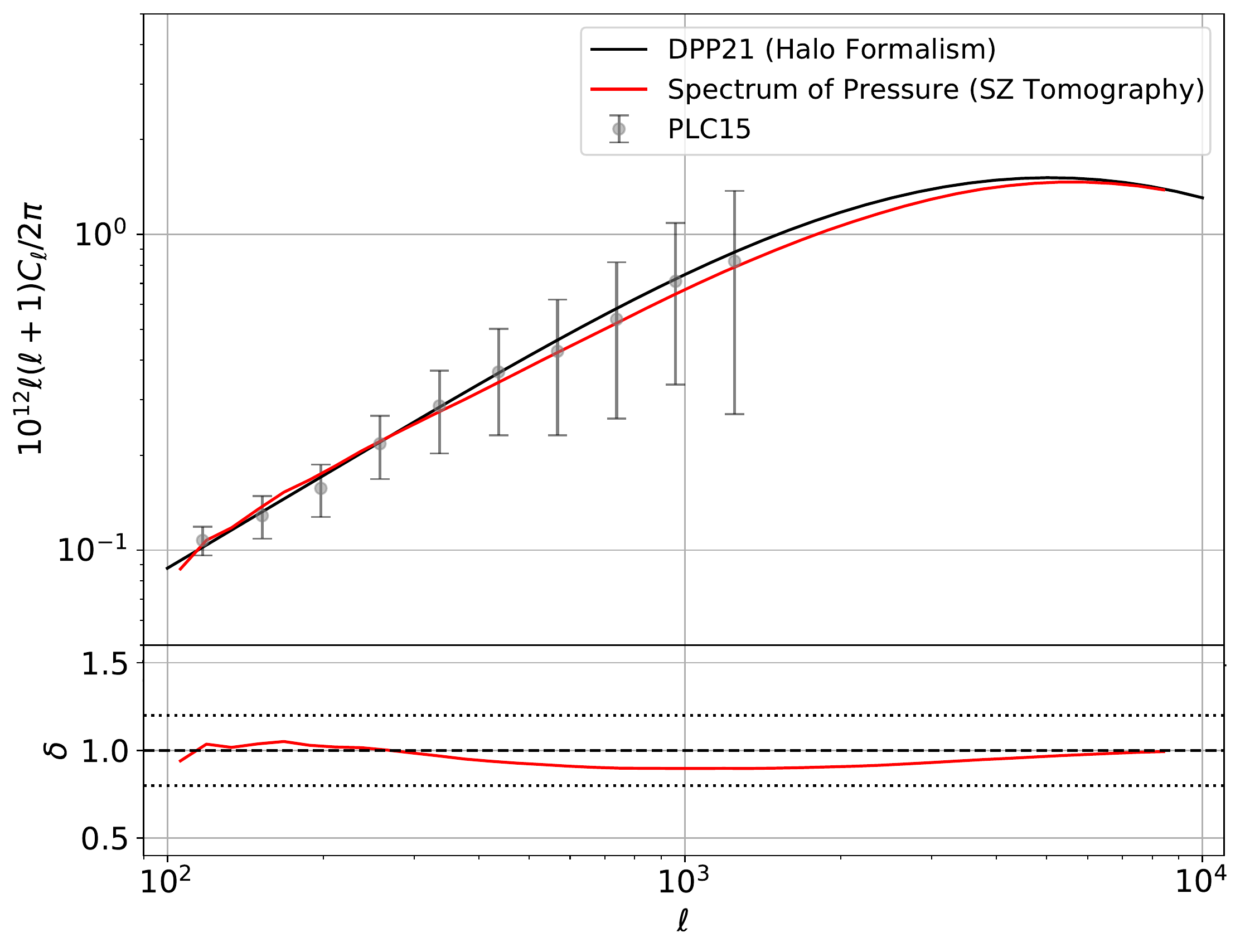}
\caption{\textbf{Top:} tSZ angular power spectrum evaluated with the 3D power
spectrum of the gas thermal pressure and Eq.~\ref{eq:tsz_p_tomo} (red) using the outputs from the HYPER simulation at different redshifts from $z=0$ to $z=5$. The HYPER result agrees within around 10\% with the predictions for the tSZ power spectrum calculated with the ICM pressure model used for constructing the HPM mapping relation, DPP \citep{2021ApJ...908...91H}, and the halo formalism (black line). Planck 2015 analysis \citep{2016A&A...594A..22P} of the tSZ power spectrum (gray dots) with error bars due to uncertainties of foreground contamination and statistical errors is also plotted for comparison. \textbf{Bottom:} Ratio of the tSZ power spectrum evaluated from the simulation outputs to the analytical prediction (red line) and $\pm$20\% region (dotted black lines).}\label{fig:tSZ_angularpower}
\end{figure}

In Figure~\ref{fig:tSZ_angularpower} we show the tSZ power spectrum measured from the HYPER simulation, which is calculated with the 3D power
spectrum of the gas thermal pressure. We compare our result to the analytical prediction evaluated with the ICM pressure model and the halo model assuming the same fiducial  cosmological parameters as adopted by simulation. We also plot the measurements of the tSZ power spectrum by Planck, which are in good agreement with the analytical calculation of the tSZ power spectrum derived from DPP in \citet{2021ApJ...908...91H}. The frequency-dependent terms are all scaled to $f^2(\nu)=1$ for direct comparison. 

The tSZ power spectrum measured with the output of the HYPER simulation is in  good agreement with the analytical prediction calculated with the ICM model, within $\sim$ 10\%. It shows that the HYPER simulation results are consistent with the assumed ICM model used to implement the hydrodynamics in our simulation not only for properties of simulated halos, but also for the most widely used statistical measures of the tSZ effect. According to Eq.~\ref{eq:tSZ}, Compton signals of galaxy clusters dominates the contribution to the tSZ power spectrum, and we show in Section~\ref{subsec:halo quantities} that scaling relation between the Compton $Y$-parameter and the mass of simulated halos matches the ICM model derivation very well, which also indicates a good agreement on the tSZ power spectrum calculation.
\section{Conclusion and Discussion} \label{sec:conclusion}

In this paper we introduce HYPER, a new implementation of a fast approximate hydro simulation based on an N-body solver. HYPER applies a power-law density-temperature relation for the gas in the IGM regime of low-density and constructs a mapping relation between two designed HPM variables and the gas temperature and pressure in the high-density ICM regime (which based on the adopted ICM gas pressure model) to simulate the evolution of baryonic matter in an efficient way.

We investigate the properties of gas inside the simulated halos by measuring the radial profiles of density, temperature, and pressure of the gas for the identified halos in the simulation. We also present the integrated quantities of observables in the X-ray and SZ survey and calculate the tSZ angular power spectrum from the simulation outputs. We emphasize that one of the crucial strength of HYPER is one can systematically control the gas physics of simulated halos with the adopted ICM model. We show that
\begin{enumerate}
    \item the radial profiles of density, temperature and pressure of gas inside the identified halos in the simulation are in good agreement with the ICM model predictions within 5$\%$ for 0.1$R_\mathrm{200c}-R_\mathrm{200c}$ and the deviation is limited to 20$\%$ in the inner core $r< 0.1R_\mathrm{200c}$ and outer skirt region $R_\mathrm{200c}-1.5R_\mathrm{200c}$. Mild inconsistency found in the inner region might be due to the resolution limit of the HPM solver.
    \item the integrated X-ray and SZ observables of simulated halos are in good agreement with the scaling relations derived from the gas radial profiles of the ICM model at redshifts from $z=0$ to $z=1$. The scatter in the relation comes from two primary sources: the variations in the dynamical states of different simulated halos and the finite numerical resolution in the inner core of cluster. The latter may also contribute to the bias of the gas fraction of  simulated halos as compared to the model prediction. The scatter in the simulation results of different observables is comparable to that in the full hydrodynamic simulation Magneticum and MACSIS.
    \item the tSZ angular power spectrum measured for the HYPER  simulation, which is calculated using the 3D power spectrum of the gas thermal pressure drawn from the simulation at different redshift snapshots, is in good agreement with the analytical predictions evaluated with the halo model and the ICM model used to implement the HPM algorithm. Good consistency in the simulation output and ICM model derivation for properties of the ICM regime includes the cluster radial profiles, SZ and X-ray observable-mass relation, and statistical quantities of the tSZ effects indicates HYPER simulation allows us to systematically control the ICM physics by varying the ICM model implemented in the HPM mapping relation construction.
\end{enumerate}

We envision three main use cases for HYPER.

\begin{enumerate}
    \item HYPER runs orders of magnitude faster than ordinary hydrodynamic simulations. It can be useful for generating a large number of mock catalogs and creating maps of various physical quantities (like X-Ray, temperature, SZ effect, etc.) for galaxy clusters. These outputs will help in the development of data reduction and analysis pipelines, for understanding systematics and selection effects, and for interpreting cosmological and astrophysical constraints. One can also envision training a multi-band deep learning model with mock observations generated by HYPER for the mass estimation and mass distribution measurement of galaxy clusters.
    \item In HYPER we implement the dynamics of baryons via the HPM mapping relation built on the ICM model for the gas profiles. Modifications of the matter power spectrum due to baryonic physics are one of the major theoretical uncertainties in cosmological weak lensing measurements. The ability of HYPER simulations to efficiently model the joint effects and varied cosmological parameters is a powerful tool for studying mitigation schemes for baryonic effects in weak lensing cosmic shear measurements \citep[e.g.][]{2019MNRAS.488.1652H}.
    
    \item Baryonic physics such as star formation, energetic feedback, and nonthermal pressure support affect the tSZ angular power spectrum in nontrivial ways \citep[e.g.][]{2011ApJ...727...94T}. The difference in gas physics is imprinted in the cluster gas pressure profile. With HYPER one can systematically vary the gas pressure model of galaxy clusters. For example, one can implement different ICM models drawn from current state-of-art high resolution hydrodynamic simulations in a large scale fast hydro simulation of size up to $\sim$ 1-2 Gpc, which would be unrealistically expensive for ordinary hydrodynamic simulations. With this approach one can systematically study how different gas physics influences, for example, the tSZ angular power spectrum. Moreover, from an inverse perspective, the efficiency of HYPER also enables us to generate a large number of mock observations, combined with proper statistic techniques (e.g. Gaussian Process) allows us to quickly explore parameter space for ICM model and use observation data of the SZ effect to put constraints on the ICM model; furthermore, it can be used to examine the gas physics implemented in cosmological simulation.
\end{enumerate}
\par
In future extensions of this work we will focus on improving the finite spatial resolution of HYPER. One avenue could be adopting an hybrid scheme combining the multigrid method with the fast Fourier transform \citep[e.g.][]{1997ApJS..111...73K}, which could eliminate the resolution effect in the high-density ICM regime and sustain the high computational efficiency throughout the rest of the simulation volume. It may help solve the problem of overestimating the gas velocity dispersion in the inner region of halos. We also plan to study the performance of HYPER compared to other large-scale state-of-art hydrodynamic simulations like Magneticum, Illustris, etc, by replacing the ICM pressure model adopted in this work with a simulated ICM pressure profile drawn from the corresponding hydrodynamic simulation. That would allow a direct comparison between HYPER and full hydro simulations, and would enable us to study further the reliability of our new HPM algorithm.
\section*{Acknowledgements}

We thank Markus Michael Rau, Qirong Zhu, and Renyue Cen for helpful discussions. HT acknowledges support from the NSF AI Institute: Physics of the Future, NSF PHY-2020295. Computer simulations and analysis were supported by Pittsburgh Supercomputing Center under grant AST190014P, PHY200019P and PHY200058P.
\appendix
\section{HPM Variables Adjustment}\label{subsec:halovar_adjust}

\begin{figure}[hbt!]
\includegraphics[width=\textwidth]{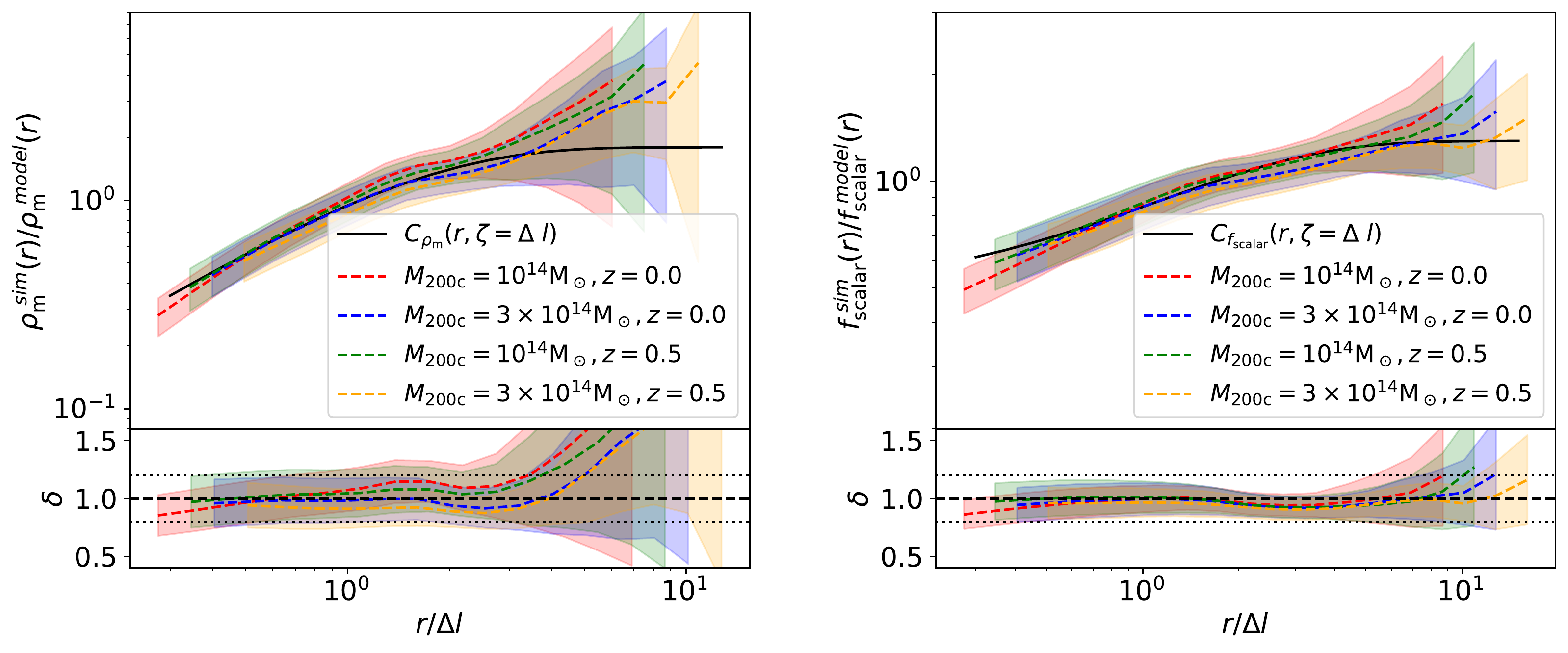}
\caption{Ratio of simulated mass-weighted profiles of the HPM variables, $\rho_{\rm m}$ (left panel) and $f_{\rm{scalar}}$ (right panel), to the theoretical derivation in the ICM model for mass bins $M_{\rm{200c}}=10^{14}{\rm M}_\odot$ and $3\times10^{14}{\rm M}_\odot$ at redshift $z=0$ and $z=0.5$ (dashed colorful lines) and their uncertainties (thin bands). Also shown are the best fits for the calibration functions $C(r,\zeta)$ (solid black line). Fitted calibration functions are found to be in good agreement with the simulation results, while deviation appears at large radii where the uncertainties for the simulated radial profiles become great more significant and data points are less important in the fitting. The bottom panels show the ratio between the simulation results and the best fits for the calibration functions and the $\pm20\%$ band (dotted black lines)}\label{fig:halovar_adjust}
\end{figure}
We adopt the form of the Weibull function for the calibration function $C(r,\zeta)$ to characterize the difference between the simulated radial profiles of HPM variables and their theoretical values in the ICM model:
\begin{align}
    C_{Y}(r,\zeta)=A_{\rm L}^C-(A_{\rm L}^C-A_{\rm H}^C)\exp[-(A_{\rm S }^C(\textit{r}/\zeta))],
\end{align}
where $Y$ denote the HPM variables $\rho_m$ and $f_{\rm scalar}$ and $\zeta=\Delta \textit{l}=\textit{L}/\textit{N}$ is the length of the grid cell in our simulation. We fit the calibration function to the ratio of the simulated mass-weighted profiles of  $\rho_{\rm m}$ and $f_{\rm scalar}$ to the radial profiles derived in the standard ICM model for two mass bins $M_{\rm{200c}}=10^{14}{\rm M}_\odot,3\times10^{14}{\rm M}_\odot$ at redshift $z=0.0,0.5$, and importance of the data points in fitting are weighted by the uncertainties of the radial profiles. The best fit results for the parameters $A_{\rm H}^C,A_{\rm L}^C,A_{\rm S}^C$ are [1.80,0.01,0.70] and [1.30,0.48,0.60] for HPM variables $\rho_{\rm m}$ and $f_{\rm scalar}$ respectively. These parameters of the calibration functions may be spatial resolution dependent.

In Figure~\ref{fig:halovar_adjust} we plot the ratio of simulated radial profiles of $\rho_{\rm m}$ and $f_{\rm scalar}$ to their respective values in the adopted ICM model and its uncertainty. We also show the best fit results for our calibration function. The fitted calibration functions agree well with the simulated mean profiles in the inner regions of halos, but show more deviation as the uncertainties of the simulation results grow in the outskirt region. 

\section{Pressure Filtering and Artificial Viscosity Tuning}\label{subsec:filter_vsc_tune}

\begin{figure}[hbt!]
\includegraphics[width=\textwidth]{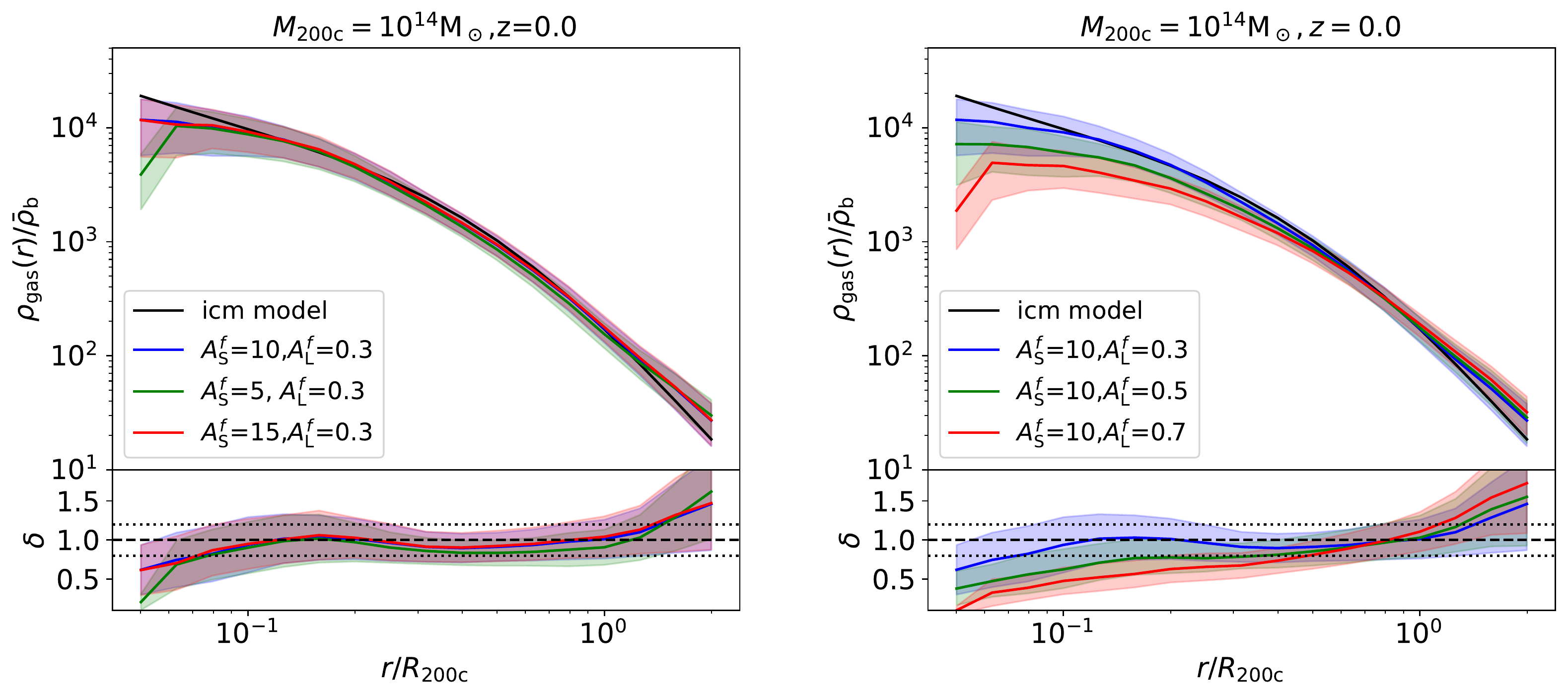}
\caption{{\bf{Left:}} Simulated gas density profiles as compared to the ICM model (solid black line) when we apply pressure filters with filter parameters set to $A_{\rm S}^f=5$ (green), $A_{\rm S}^f=10$ (blue), and $A_{\rm S}^f=15$ (red). Simulation results for the gas density profiles are insensitive to the pressure filter parameter $A_{\rm S}^f$. Ratio between simulated gas density profile and the ICM model and the $\pm$ 20\% region are shown in the bottom panel. {\bf{Right:}} Simulated gas density profiles  with filter parameters $A_{\rm L}^f=0.3$ (blue), $A_{\rm L}^f=0.5$ (green), and $A_{\rm L}^f=0.7$ (red) compare to the ICM model (black).}\label{fig:filter_tune}
\end{figure}
 As discussed in Section~\ref{subsec:filter_vsc}, we apply a pressure filter on the gas pressure field in Fourier space to smooth the gas pressure in the inner region of the halos and suppress the numerical noise. The pressure filter is in the form of the Weibull function expressed by Eq.~\ref{eq:p_filter}. We tune the parameters $A_{\rm L}$ and $A_{\rm S}$, which determine the scale range and the degree of smoothing at small scales for the pressure filter, to make the simulated gas density profiles match the ICM profiles.

In Figure~\ref{fig:filter_tune} we compare the simulated gas density profiles with the analytical derivation in the ICM model after filtering the gas pressure field in Fourier space. We experiment with using the pressure filters with different combinations of $A_{\rm S}^f$ and $A_{\rm L}^f$. We find that the simulation results are less sensitive to the filter parameter $A_{\rm S}^f$, and we choose to adopt $A_{\rm S}^f=10$ for our simulation. We tune the $A_{\rm L}^f$ starting from $A_{\rm L}^f=1.0$ then gradually decrease to adjust the smoothing effect on the gas pressure imposed by the filter until we can balance the hydro force and the gravity and get a proper density profile when $A_{\rm L}^f=0.3$.  
\begin{figure}[hbt!]
\includegraphics[width=\textwidth]{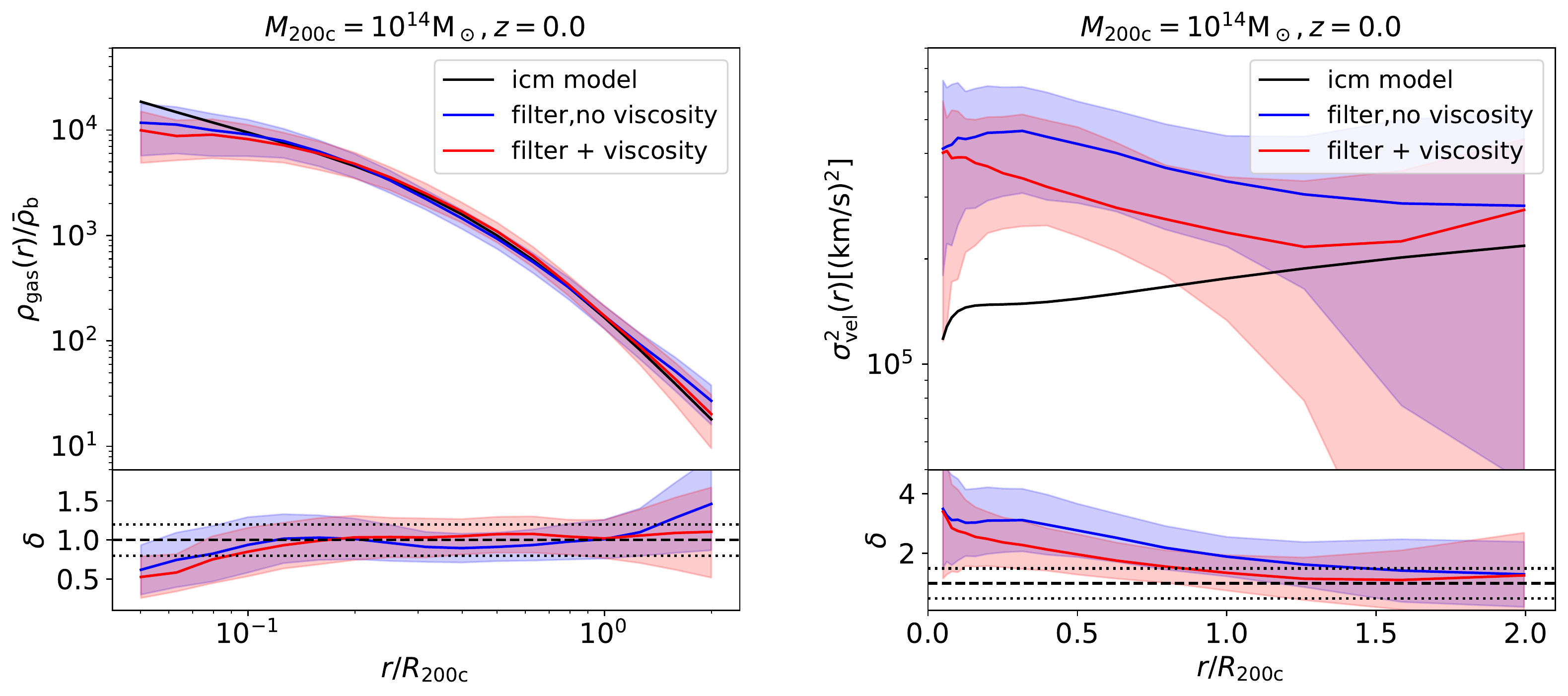}
\caption{Simulated gas density profile (left) and gas velocity dispersion (right) and their uncertainties with both pressure filter and artificial viscosity applied in the simulation (red) as compared to the case with only the pressure filter being applied (blue). Both cases show good agreements with the ICM model (solid black line). Their ratio and the $\pm$ 20\% (left), $\pm$ 50\% (right) region (dotted black lines) are shown in the bottom panel.} \label{fig:filter_vsc_tune}
\end{figure}

However, with only the pressure filter, we can obtain a proper density profile but overestimate the velocity dispersion for gas particles in the simulated halos. By applying the artificial viscosity, which removes the extra kinetic energy of gas particles that should have been converted into heat, we can reduce the discrepancy, although not eliminate it completely. As shown in the right panel of Figure~\ref{fig:filter_tune}, smoothing the gas thermal pressure field avoids an excessive hydro force in the inner region of the simulated halos and prevents pushing too much gas out of the halo. We also find that the artificial viscosity can effectively reduce the gas velocity dispersion and decreases the non-thermal pressure. Since both pressure filter and artificial viscosity can mitigate the gas pressure to balance the hydro force and gravity, we can tune the pressure filter and the artificial viscosity jointly to keep the simulated gas density profile matching the model derivation and reduce the gas velocity dispersion at the same time. In this case, we find the pressure filter with slightly larger $A_{\rm L}^f$ that $A_{\rm L}^f=0.5$ and $\alpha\sim 0.1, \beta\sim 0.05$ for the parameters of artificial viscosity in Eq.~\ref{eq:artificial_viscosity_Q} work best for our simulation.

In Figure~\ref{fig:filter_vsc_tune} we plot the ICM gas profiles with (a) both pressure filter and artificial viscosity and (b) the pressure filter only. We show that by tuning the pressure filtering and the artificial viscosity jointly, we can make the gas density profile match the ICM model and obtain a lower gas velocity dispersion in the halo outskirt region. 

 Even with the artificial viscosity included into our simulation, it is still difficult to match the gas velocity dispersion in the inner region of the simulated halos to the ICM model. This may be because the finite resolution of the HPM algorithm limits the ability to accurately resolve the local artificial viscosity exerted on the gas particles in our simulation,  especially in the high-density halo center region where hundreds even thousands of particles can co-locate in one grid cell. Further studies are required to explore the reason for the discrepancy and improve the simulation fidelity. One possible solution to increase the resolution in high-density regions is to use the adaptive mesh refinement.

\bibliography{hyper}{}

\begin{thebibliography}{}
\expandafter\ifx\csname natexlab\endcsname\relax\def\natexlab#1{#1}\fi
\providecommand{\url}[1]{\href{#1}{#1}}
\providecommand{\dodoi}[1]{doi:~\href{http://doi.org/#1}{\nolinkurl{#1}}}
\providecommand{\doeprint}[1]{\href{http://ascl.net/#1}{\nolinkurl{http://ascl.net/#1}}}
\providecommand{\doarXiv}[1]{\href{https://arxiv.org/abs/#1}{\nolinkurl{https://arxiv.org/abs/#1}}}

\bibitem[{{Angulo} {et~al.}(2012){Angulo}, {Springel}, {White}, {Jenkins},
  {Baugh}, \& {Frenk}}]{2012MNRAS.426.2046A}
{Angulo}, R.~E., {Springel}, V., {White}, S.~D.~M., {et~al.} 2012, \mnras, 426,
  2046, \dodoi{10.1111/j.1365-2966.2012.21830.x}

\bibitem[{{Arnaud} {et~al.}(2010){Arnaud}, {Pratt}, {Piffaretti},
  {B{\"o}hringer}, {Croston}, \& {Pointecouteau}}]{2010A&A...517A..92A}
{Arnaud}, M., {Pratt}, G.~W., {Piffaretti}, R., {et~al.} 2010, \aap, 517, A92,
  \dodoi{10.1051/0004-6361/200913416}

\bibitem[{{Barnes} {et~al.}(2017){Barnes}, {Kay}, {Henson}, {McCarthy},
  {Schaye}, \& {Jenkins}}]{2017MNRAS.465..213B}
{Barnes}, D.~J., {Kay}, S.~T., {Henson}, M.~A., {et~al.} 2017, \mnras, 465,
  213, \dodoi{10.1093/mnras/stw2722}

\bibitem[{{Battaglia} {et~al.}(2012{\natexlab{a}}){Battaglia}, {Bond},
  {Pfrommer}, \& {Sievers}}]{2012ApJ...758...75B}
{Battaglia}, N., {Bond}, J.~R., {Pfrommer}, C., \& {Sievers}, J.~L.
  2012{\natexlab{a}}, \apj, 758, 75, \dodoi{10.1088/0004-637X/758/2/75}

\bibitem[{{Battaglia} {et~al.}(2012{\natexlab{b}}){Battaglia}, {Bond},
  {Pfrommer}, \& {Sievers}}]{2012ApJ...758...74B}
---. 2012{\natexlab{b}}, \apj, 758, 74, \dodoi{10.1088/0004-637X/758/2/74}

\bibitem[{{Bautz} {et~al.}(2009){Bautz}, {Miller}, {Sanders}, {Arnaud},
  {Mushotzky}, {Porter}, {Hayashida}, {Henry}, {Hughes}, {Kawaharada},
  {Makashima}, {Sato}, \& {Tamura}}]{2009PASJ...61.1117B}
{Bautz}, M.~W., {Miller}, E.~D., {Sanders}, J.~S., {et~al.} 2009, \pasj, 61,
  1117, \dodoi{10.1093/pasj/61.5.1117}

\bibitem[{{Bocquet} {et~al.}(2016){Bocquet}, {Saro}, {Dolag}, \&
  {Mohr}}]{2016MNRAS.456.2361B}
{Bocquet}, S., {Saro}, A., {Dolag}, K., \& {Mohr}, J.~J. 2016, \mnras, 456,
  2361, \dodoi{10.1093/mnras/stv2657}

\bibitem[{{Boylan-Kolchin} {et~al.}(2009){Boylan-Kolchin}, {Springel}, {White},
  {Jenkins}, \& {Lemson}}]{2009MNRAS.398.1150B}
{Boylan-Kolchin}, M., {Springel}, V., {White}, S. D.~M., {Jenkins}, A., \&
  {Lemson}, G. 2009, \mnras, 398, 1150,
  \dodoi{10.1111/j.1365-2966.2009.15191.x}

\bibitem[{{Cole} \& {Kaiser}(1988)}]{1988MNRAS.233..637C}
{Cole}, S., \& {Kaiser}, N. 1988, \mnras, 233, 637,
  \dodoi{10.1093/mnras/233.3.637}

\bibitem[{{Dav{\'e}} {et~al.}(2019){Dav{\'e}}, {Angl{\'e}s-Alc{\'a}zar},
  {Narayanan}, {Li}, {Rafieferantsoa}, \& {Appleby}}]{2019MNRAS.486.2827D}
{Dav{\'e}}, R., {Angl{\'e}s-Alc{\'a}zar}, D., {Narayanan}, D., {et~al.} 2019,
  \mnras, 486, 2827, \dodoi{10.1093/mnras/stz937}

\bibitem[{{Diemer} \& {Kravtsov}(2015)}]{2015ApJ...799..108D}
{Diemer}, B., \& {Kravtsov}, A.~V. 2015, \apj, 799, 108,
  \dodoi{10.1088/0004-637X/799/1/108}

\bibitem[{{Dolag} {et~al.}(2016){Dolag}, {Komatsu}, \&
  {Sunyaev}}]{2016MNRAS.463.1797D}
{Dolag}, K., {Komatsu}, E., \& {Sunyaev}, R. 2016, \mnras, 463, 1797,
  \dodoi{10.1093/mnras/stw2035}

\bibitem[{{Doroshkevich} {et~al.}(1980){Doroshkevich}, {Kotok}, {Poliudov},
  {Shandarin}, {Sigov}, \& {Novikov}}]{1980MNRAS.192..321D}
{Doroshkevich}, A.~G., {Kotok}, E.~V., {Poliudov}, A.~N., {et~al.} 1980,
  \mnras, 192, 321, \dodoi{10.1093/mnras/192.2.321}

\bibitem[{{Dubois} {et~al.}(2016){Dubois}, {Peirani}, {Pichon}, {Devriendt},
  {Gavazzi}, {Welker}, \& {Volonteri}}]{2016MNRAS.463.3948D}
{Dubois}, Y., {Peirani}, S., {Pichon}, C., {et~al.} 2016, \mnras, 463, 3948,
  \dodoi{10.1093/mnras/stw2265}

\bibitem[{{Feng} {et~al.}(2016{\natexlab{a}}){Feng}, {Chu}, {Seljak}, \&
  {McDonald}}]{2016MNRAS.463.2273F}
{Feng}, Y., {Chu}, M.-Y., {Seljak}, U., \& {McDonald}, P. 2016{\natexlab{a}},
  \mnras, 463, 2273, \dodoi{10.1093/mnras/stw2123}

\bibitem[{{Feng} {et~al.}(2016{\natexlab{b}}){Feng}, {Di-Matteo}, {Croft},
  {Bird}, {Battaglia}, \& {Wilkins}}]{2016MNRAS.455.2778F}
{Feng}, Y., {Di-Matteo}, T., {Croft}, R.~A., {et~al.} 2016{\natexlab{b}},
  \mnras, 455, 2778, \dodoi{10.1093/mnras/stv2484}

\bibitem[{{George} {et~al.}(2009){George}, {Fabian}, {Sanders}, {Young}, \&
  {Russell}}]{2009MNRAS.395..657G}
{George}, M.~R., {Fabian}, A.~C., {Sanders}, J.~S., {Young}, A.~J., \&
  {Russell}, H.~R. 2009, \mnras, 395, 657,
  \dodoi{10.1111/j.1365-2966.2009.14547.x}

\bibitem[{{Gnedin}(1995)}]{1995ApJS...97..231G}
{Gnedin}, N.~Y. 1995, \apjs, 97, 231, \dodoi{10.1086/192141}

\bibitem[{{Gnedin} \& {Hui}(1998)}]{1998MNRAS.296...44G}
{Gnedin}, N.~Y., \& {Hui}, L. 1998, \mnras, 296, 44,
  \dodoi{10.1046/j.1365-8711.1998.01249.x}

\bibitem[{{Gupta} {et~al.}(2017){Gupta}, {Saro}, {Mohr}, {Dolag}, \&
  {Liu}}]{2017MNRAS.469.3069G}
{Gupta}, N., {Saro}, A., {Mohr}, J.~J., {Dolag}, K., \& {Liu}, J. 2017, \mnras,
  469, 3069, \dodoi{10.1093/mnras/stx715}

\bibitem[{{He} {et~al.}(2021){He}, {Mansfield}, {Rau}, {Trac}, \&
  {Battaglia}}]{2021ApJ...908...91H}
{He}, Y., {Mansfield}, P., {Rau}, M.~M., {Trac}, H., \& {Battaglia}, N. 2021,
  \apj, 908, 91, \dodoi{10.3847/1538-4357/abd0ff}

\bibitem[{{Hockney} \& {Eastwood}(1981)}]{1981csup.book.....H}
{Hockney}, R.~W., \& {Eastwood}, J.~W. 1981, {Computer Simulation Using
  Particles}

\bibitem[{{Hoshino} {et~al.}(2010){Hoshino}, {Henry}, {Sato}, {Akamatsu},
  {Yokota}, {Sasaki}, {Ishisaki}, {Ohashi}, {Bautz}, {Fukazawa}, {Kawano},
  {Furuzawa}, {Hayashida}, {Tawa}, {Hughes}, {Kokubun}, \&
  {Tamura}}]{2010PASJ...62..371H}
{Hoshino}, A., {Henry}, J.~P., {Sato}, K., {et~al.} 2010, \pasj, 62, 371,
  \dodoi{10.1093/pasj/62.2.371}

\bibitem[{{Huang} {et~al.}(2019){Huang}, {Eifler}, {Mandelbaum}, \&
  {Dodelson}}]{2019MNRAS.488.1652H}
{Huang}, H.-J., {Eifler}, T., {Mandelbaum}, R., \& {Dodelson}, S. 2019, \mnras,
  488, 1652, \dodoi{10.1093/mnras/stz1714}

\bibitem[{{Hui} \& {Gnedin}(1997)}]{1997MNRAS.292...27H}
{Hui}, L., \& {Gnedin}, N.~Y. 1997, \mnras, 292, 27,
  \dodoi{10.1093/mnras/292.1.27}

\bibitem[{{Iannuzzi} \& {Dolag}(2012)}]{2012MNRAS.427.1024I}
{Iannuzzi}, F., \& {Dolag}, K. 2012, \mnras, 427, 1024,
  \dodoi{10.1111/j.1365-2966.2012.22017.x}

\bibitem[{{Ishiyama} {et~al.}(2021){Ishiyama}, {Prada}, {Klypin}, {Sinha},
  {Metcalf}, {Jullo}, {Altieri}, {Cora}, {Croton}, {de la Torre},
  {Mill{\'a}n-Calero}, {Oogi}, {Ruedas}, \&
  {Vega-Mart{\'\i}nez}}]{2021MNRAS.tmp.1529I}
{Ishiyama}, T., {Prada}, F., {Klypin}, A.~A., {et~al.} 2021, \mnras,
  \dodoi{10.1093/mnras/stab1755}

\bibitem[{{Kaiser}(1986)}]{1986MNRAS.222..323K}
{Kaiser}, N. 1986, \mnras, 222, 323, \dodoi{10.1093/mnras/222.2.323}

\bibitem[{{Kaviraj} {et~al.}(2017){Kaviraj}, {Laigle}, {Kimm}, {Devriendt},
  {Dubois}, {Pichon}, {Slyz}, {Chisari}, \& {Peirani}}]{2017MNRAS.467.4739K}
{Kaviraj}, S., {Laigle}, C., {Kimm}, T., {et~al.} 2017, \mnras, 467, 4739,
  \dodoi{10.1093/mnras/stx126}

\bibitem[{{Kawaharada} {et~al.}(2010){Kawaharada}, {Okabe}, {Umetsu},
  {Takizawa}, {Matsushita}, {Fukazawa}, {Hamana}, {Miyazaki}, {Nakazawa}, \&
  {Ohashi}}]{2010ApJ...714..423K}
{Kawaharada}, M., {Okabe}, N., {Umetsu}, K., {et~al.} 2010, \apj, 714, 423,
  \dodoi{10.1088/0004-637X/714/1/423}

\bibitem[{{Khandai} {et~al.}(2015){Khandai}, {Di Matteo}, {Croft}, {Wilkins},
  {Feng}, {Tucker}, {DeGraf}, \& {Liu}}]{2015MNRAS.450.1349K}
{Khandai}, N., {Di Matteo}, T., {Croft}, R., {et~al.} 2015, \mnras, 450, 1349,
  \dodoi{10.1093/mnras/stv627}

\bibitem[{{Klypin} {et~al.}(2016){Klypin}, {Yepes}, {Gottl{\"o}ber}, {Prada},
  \& {He{\ss}}}]{2016MNRAS.457.4340K}
{Klypin}, A., {Yepes}, G., {Gottl{\"o}ber}, S., {Prada}, F., \& {He{\ss}}, S.
  2016, \mnras, 457, 4340, \dodoi{10.1093/mnras/stw248}

\bibitem[{{Klypin} {et~al.}(2011){Klypin}, {Trujillo-Gomez}, \&
  {Primack}}]{2011ApJ...740..102K}
{Klypin}, A.~A., {Trujillo-Gomez}, S., \& {Primack}, J. 2011, \apj, 740, 102,
  \dodoi{10.1088/0004-637X/740/2/102}

\bibitem[{Komatsu \& Kitayama(1999)}]{1538-4357-526-1-L1}
Komatsu, E., \& Kitayama, T. 1999, The Astrophysical Journal Letters, 526, L1.
\newblock \url{http://stacks.iop.org/1538-4357/526/i=1/a=L1}

\bibitem[{{Komatsu} \& {Seljak}(2002)}]{2002MNRAS.336.1256K}
{Komatsu}, E., \& {Seljak}, U. 2002, \mnras, 336, 1256,
  \dodoi{10.1046/j.1365-8711.2002.05889.x}

\bibitem[{{Kravtsov} {et~al.}(1997){Kravtsov}, {Klypin}, \&
  {Khokhlov}}]{1997ApJS..111...73K}
{Kravtsov}, A.~V., {Klypin}, A.~A., \& {Khokhlov}, A.~M. 1997, \apjs, 111, 73,
  \dodoi{10.1086/313015}

\bibitem[{{Kravtsov} {et~al.}(2006){Kravtsov}, {Vikhlinin}, \&
  {Nagai}}]{2006ApJ...650..128K}
{Kravtsov}, A.~V., {Vikhlinin}, A., \& {Nagai}, D. 2006, \apj, 650, 128,
  \dodoi{10.1086/506319}

\bibitem[{{Lau} {et~al.}(2013){Lau}, {Nagai}, \&
  {Nelson}}]{2013ApJ...777..151L}
{Lau}, E.~T., {Nagai}, D., \& {Nelson}, K. 2013, \apj, 777, 151,
  \dodoi{10.1088/0004-637X/777/2/151}

\bibitem[{{Limber}(1953)}]{1953ApJ...117..134L}
{Limber}, D.~N. 1953, \apj, 117, 134, \dodoi{10.1086/145672}

\bibitem[{{Mandelbaum} {et~al.}(2006){Mandelbaum}, {Seljak}, {Cool}, {Blanton},
  {Hirata}, \& {Brinkmann}}]{2006MNRAS.372..758M}
{Mandelbaum}, R., {Seljak}, U., {Cool}, R.~J., {et~al.} 2006, \mnras, 372, 758,
  \dodoi{10.1111/j.1365-2966.2006.10906.x}

\bibitem[{{Martel} \& {Shapiro}(1998)}]{1998MNRAS.297..467M}
{Martel}, H., \& {Shapiro}, P.~R. 1998, \mnras, 297, 467,
  \dodoi{10.1046/j.1365-8711.1998.01497.x}

\bibitem[{{McCarthy} {et~al.}(2017){McCarthy}, {Schaye}, {Bird}, \& {Le
  Brun}}]{2017MNRAS.465.2936M}
{McCarthy}, I.~G., {Schaye}, J., {Bird}, S., \& {Le Brun}, A. M.~C. 2017,
  \mnras, 465, 2936, \dodoi{10.1093/mnras/stw2792}

\bibitem[{{McDonald} {et~al.}(2002){McDonald}, {Miralda-Escud{\'e}}, \&
  {Cen}}]{2002ApJ...580...42M}
{McDonald}, P., {Miralda-Escud{\'e}}, J., \& {Cen}, R. 2002, \apj, 580, 42,
  \dodoi{10.1086/343031}

\bibitem[{{Monaco} {et~al.}(2002){Monaco}, {Theuns}, \&
  {Taffoni}}]{2002MNRAS.331..587M}
{Monaco}, P., {Theuns}, T., \& {Taffoni}, G. 2002, \mnras, 331, 587,
  \dodoi{10.1046/j.1365-8711.2002.05162.x}

\bibitem[{{Morandi} \& {Limousin}(2012)}]{2012MNRAS.421.3147M}
{Morandi}, A., \& {Limousin}, M. 2012, \mnras, 421, 3147,
  \dodoi{10.1111/j.1365-2966.2012.20537.x}

\bibitem[{{Nagai} {et~al.}(2007){Nagai}, {Kravtsov}, \&
  {Vikhlinin}}]{2007ApJ...668....1N}
{Nagai}, D., {Kravtsov}, A.~V., \& {Vikhlinin}, A. 2007, \apj, 668, 1,
  \dodoi{10.1086/521328}

\bibitem[{{Navarro} {et~al.}(1996){Navarro}, {Frenk}, \&
  {White}}]{1996ApJ...462..563N}
{Navarro}, J.~F., {Frenk}, C.~S., \& {White}, S. D.~M. 1996, \apj, 462, 563,
  \dodoi{10.1086/177173}

\bibitem[{{Navarro} {et~al.}(1997){Navarro}, {Frenk}, \&
  {White}}]{1997ApJ...490..493N}
---. 1997, \apj, 490, 493, \dodoi{10.1086/304888}

\bibitem[{{Navarro} {et~al.}(2004){Navarro}, {Hayashi}, {Power}, {Jenkins},
  {Frenk}, {White}, {Springel}, {Stadel}, \& {Quinn}}]{2004MNRAS.349.1039N}
{Navarro}, J.~F., {Hayashi}, E., {Power}, C., {et~al.} 2004, \mnras, 349, 1039,
  \dodoi{10.1111/j.1365-2966.2004.07586.x}

\bibitem[{Nelson {et~al.}(2014)Nelson, Lau, \& Nagai}]{0004-637X-792-1-25}
Nelson, K., Lau, E.~T., \& Nagai, D. 2014, The Astrophysical Journal, 792, 25.
\newblock \url{http://stacks.iop.org/0004-637X/792/i=1/a=25}

\bibitem[{{Nelson} {et~al.}(2012){Nelson}, {Rudd}, {Shaw}, \&
  {Nagai}}]{2012ApJ...751..121N}
{Nelson}, K., {Rudd}, D.~H., {Shaw}, L., \& {Nagai}, D. 2012, \apj, 751, 121,
  \dodoi{10.1088/0004-637X/751/2/121}

\bibitem[{{Olamaie} {et~al.}(2012){Olamaie}, {Hobson}, \&
  {Grainge}}]{2012MNRAS.423.1534O}
{Olamaie}, M., {Hobson}, M.~P., \& {Grainge}, K. J.~B. 2012, \mnras, 423, 1534,
  \dodoi{10.1111/j.1365-2966.2012.20980.x}

\bibitem[{{Pen}(1998)}]{1998ApJS..115...19P}
{Pen}, U.-L. 1998, \apjs, 115, 19, \dodoi{10.1086/313074}

\bibitem[{{Pillepich} {et~al.}(2018){Pillepich}, {Springel}, {Nelson}, {Genel},
  {Naiman}, {Pakmor}, {Hernquist}, {Torrey}, {Vogelsberger}, {Weinberger}, \&
  {Marinacci}}]{2018MNRAS.473.4077P}
{Pillepich}, A., {Springel}, V., {Nelson}, D., {et~al.} 2018, \mnras, 473,
  4077, \dodoi{10.1093/mnras/stx2656}

\bibitem[{{Planck Collaboration} {et~al.}(2016){Planck Collaboration},
  {Aghanim}, {Arnaud}, {Ashdown}, {Aumont}, {Baccigalupi}, {Banday},
  {Barreiro}, {Bartlett}, {Bartolo}, \& et~al.}]{2016A&A...594A..22P}
{Planck Collaboration}, {Aghanim}, N., {Arnaud}, M., {et~al.} 2016, \aap, 594,
  A22, \dodoi{10.1051/0004-6361/201525826}

\bibitem[{{Reiprich} {et~al.}(2009){Reiprich}, {Hudson}, {Zhang}, {Sato},
  {Ishisaki}, {Hoshino}, {Ohashi}, {Ota}, \& {Fujita}}]{2009A&A...501..899R}
{Reiprich}, T.~H., {Hudson}, D.~S., {Zhang}, Y.-Y., {et~al.} 2009, \aap, 501,
  899, \dodoi{10.1051/0004-6361/200810404}

\bibitem[{{Schaye} {et~al.}(2015){Schaye}, {Crain}, {Bower}, {Furlong},
  {Schaller}, {Theuns}, {Dalla Vecchia}, {Frenk}, {McCarthy}, {Helly},
  {Jenkins}, {Rosas-Guevara}, {White}, {Baes}, {Booth}, {Camps}, {Navarro},
  {Qu}, {Rahmati}, {Sawala}, {Thomas}, \& {Trayford}}]{2015MNRAS.446..521S}
{Schaye}, J., {Crain}, R.~A., {Bower}, R.~G., {et~al.} 2015, \mnras, 446, 521,
  \dodoi{10.1093/mnras/stu2058}

\bibitem[{{Schmidt} \& {Allen}(2007)}]{2007MNRAS.379..209S}
{Schmidt}, R.~W., \& {Allen}, S.~W. 2007, \mnras, 379, 209,
  \dodoi{10.1111/j.1365-2966.2007.11928.x}

\bibitem[{{Scoccimarro} \& {Sheth}(2002)}]{2002MNRAS.329..629S}
{Scoccimarro}, R., \& {Sheth}, R.~K. 2002, \mnras, 329, 629,
  \dodoi{10.1046/j.1365-8711.2002.04999.x}

\bibitem[{Shaw {et~al.}(2010)Shaw, Nagai, Bhattacharya, \&
  Lau}]{0004-637X-725-2-1452}
Shaw, L.~D., Nagai, D., Bhattacharya, S., \& Lau, E.~T. 2010, The Astrophysical
  Journal, 725, 1452.
\newblock \url{http://stacks.iop.org/0004-637X/725/i=2/a=1452}

\bibitem[{{Shi} \& {Komatsu}(2014)}]{2014MNRAS.442..521S}
{Shi}, X., \& {Komatsu}, E. 2014, \mnras, 442, 521,
  \dodoi{10.1093/mnras/stu858}

\bibitem[{{Simionescu} {et~al.}(2011){Simionescu}, {Allen}, {Mantz}, {Werner},
  {Takei}, {Morris}, {Fabian}, {Sanders}, {Nulsen}, {George}, \&
  {Taylor}}]{2011Sci...331.1576S}
{Simionescu}, A., {Allen}, S.~W., {Mantz}, A., {et~al.} 2011, Science, 331,
  1576, \dodoi{10.1126/science.1200331}

\bibitem[{{Skillman} {et~al.}(2014){Skillman}, {Warren}, {Turk}, {Wechsler},
  {Holz}, \& {Sutter}}]{2014arXiv1407.2600S}
{Skillman}, S.~W., {Warren}, M.~S., {Turk}, M.~J., {et~al.} 2014, arXiv
  e-prints, arXiv:1407.2600.
\newblock \doarXiv{1407.2600}

\bibitem[{{Springel}(2005)}]{2005MNRAS.364.1105S}
{Springel}, V. 2005, \mnras, 364, 1105,
  \dodoi{10.1111/j.1365-2966.2005.09655.x}

\bibitem[{{Springel} {et~al.}(2006){Springel}, {Frenk}, \&
  {White}}]{2006Natur.440.1137S}
{Springel}, V., {Frenk}, C.~S., \& {White}, S. D.~M. 2006, \nat, 440, 1137,
  \dodoi{10.1038/nature04805}

\bibitem[{{Springel} {et~al.}(2005){Springel}, {White}, {Jenkins}, {Frenk},
  {Yoshida}, {Gao}, {Navarro}, {Thacker}, {Croton}, {Helly}, {Peacock}, {Cole},
  {Thomas}, {Couchman}, {Evrard}, {Colberg}, \& {Pearce}}]{2005Natur.435..629S}
{Springel}, V., {White}, S. D.~M., {Jenkins}, A., {et~al.} 2005, \nat, 435,
  629, \dodoi{10.1038/nature03597}

\bibitem[{{Tassev} {et~al.}(2013){Tassev}, {Zaldarriaga}, \&
  {Eisenstein}}]{2013JCAP...06..036T}
{Tassev}, S., {Zaldarriaga}, M., \& {Eisenstein}, D.~J. 2013, \jcap, 2013, 036,
  \dodoi{10.1088/1475-7516/2013/06/036}

\bibitem[{{Tinker} {et~al.}(2008){Tinker}, {Kravtsov}, {Klypin}, {Abazajian},
  {Warren}, {Yepes}, {Gottl{\"o}ber}, \& {Holz}}]{2008ApJ...688..709T}
{Tinker}, J., {Kravtsov}, A.~V., {Klypin}, A., {et~al.} 2008, \apj, 688, 709,
  \dodoi{10.1086/591439}

\bibitem[{{Trac} {et~al.}(2011){Trac}, {Bode}, \&
  {Ostriker}}]{2011ApJ...727...94T}
{Trac}, H., {Bode}, P., \& {Ostriker}, J.~P. 2011, \apj, 727, 94,
  \dodoi{10.1088/0004-637X/727/2/94}

\bibitem[{{Trac} \& {Pen}(2004)}]{2004NewA....9..443T}
{Trac}, H., \& {Pen}, U.-L. 2004, \na, 9, 443,
  \dodoi{10.1016/j.newast.2004.02.002}

\bibitem[{{Tremmel} {et~al.}(2017){Tremmel}, {Karcher}, {Governato},
  {Volonteri}, {Quinn}, {Pontzen}, {Anderson}, \&
  {Bellovary}}]{2017MNRAS.470.1121T}
{Tremmel}, M., {Karcher}, M., {Governato}, F., {et~al.} 2017, \mnras, 470,
  1121, \dodoi{10.1093/mnras/stx1160}

\bibitem[{{Truelove} {et~al.}(1997){Truelove}, {Klein}, {McKee}, {Holliman},
  {Howell}, \& {Greenough}}]{1997ApJ...489L.179T}
{Truelove}, J.~K., {Klein}, R.~I., {McKee}, C.~F., {et~al.} 1997, \apjl, 489,
  L179, \dodoi{10.1086/310975}

\bibitem[{Urban {et~al.}(2011)Urban, Werner, Simionescu, Allen, \&
  B?hringer}]{doi:10.1111/j.1365-2966.2011.18526.x}
Urban, O., Werner, N., Simionescu, A., Allen, S.~W., \& B?hringer, H. 2011,
  Monthly Notices of the Royal Astronomical Society, 414, 2101,
  \dodoi{10.1111/j.1365-2966.2011.18526.x}

\bibitem[{{Vogelsberger} {et~al.}(2020){Vogelsberger}, {Marinacci}, {Torrey},
  \& {Puchwein}}]{2020NatRP...2...42V}
{Vogelsberger}, M., {Marinacci}, F., {Torrey}, P., \& {Puchwein}, E. 2020,
  Nature Reviews Physics, 2, 42, \dodoi{10.1038/s42254-019-0127-2}

\bibitem[{{Vogelsberger} {et~al.}(2014){Vogelsberger}, {Genel}, {Springel},
  {Torrey}, {Sijacki}, {Xu}, {Snyder}, {Bird}, {Nelson}, \&
  {Hernquist}}]{2014Natur.509..177V}
{Vogelsberger}, M., {Genel}, S., {Springel}, V., {et~al.} 2014, \nat, 509, 177,
  \dodoi{10.1038/nature13316}

\bibitem[{{Voit}(2005)}]{2005RvMP...77..207V}
{Voit}, G.~M. 2005, Reviews of Modern Physics, 77, 207,
  \dodoi{10.1103/RevModPhys.77.207}

\bibitem[{{Watson} {et~al.}(2013){Watson}, {Iliev}, {D'Aloisio}, {Knebe},
  {Shapiro}, \& {Yepes}}]{2013MNRAS.433.1230W}
{Watson}, W.~A., {Iliev}, I.~T., {D'Aloisio}, A., {et~al.} 2013, \mnras, 433,
  1230, \dodoi{10.1093/mnras/stt791}

\end{thebibliography}
\bibliographystyle{aasjournal}

\end{document}